\RequirePackage{fix-cm}
\documentclass[smallextended]{svjour3}       
\smartqed  
\usepackage{graphicx}

\usepackage{mathptmx}      
\usepackage{amssymb}
\usepackage{amsmath}
\usepackage{breqn}
\usepackage{url}
\usepackage{graphicx}
\usepackage{subfig}
\usepackage{mathtools}
\usepackage[T1]{fontenc}

\begin{document}
	
	\title{An Application of Network Lasso Optimization \\for Ride Sharing Prediction 
	}
	
	
	\author{Shaona Ghosh \and Kevin Page \and
		David De Roure 
	}
	
	
	\institute{Shaona Ghosh \at
		University of Oxford, UK \\
		\email{shaona.ghosh@oerc.ox.ac.uk}           
		\and
		Kevin Page \at
		University of Oxford, UK \\
		\email{kevin.page@oerc.ox.ac.uk}
		\and
		David De Roure \at
		University of Oxford, UK \\
		\email{david.deroure@oerc.ox.ac.uk}
	}
	
	\date{}

	\maketitle
	
	\begin{abstract}
		Ride sharing has important implications in terms of environmental, social and individual goals by reducing carbon footprints, fostering social interactions and economizing commuter costs.
		The ride sharing systems that are commonly available lack adaptive and scalable techniques that can simultaneously learn from the large scale data and predict in real-time dynamic fashion. In this paper, we study such a problem towards a smart city initiative, where a generic ride sharing system is conceived capable of making predictions about ride share opportunities based on the historically recorded data while satisfying real-time ride requests. Underpinning the system is an application of a powerful machine learning convex optimization framework called Network Lasso that uses the Alternate Direction Method of Multipliers (ADMM) optimization for learning and dynamic prediction. We propose an application of a robust and scalable unified optimization framework within the ride sharing case-study. The application of Network Lasso framework is capable of jointly optimizing and clustering different rides based on their spatial and model similarity. The prediction from the framework clusters new ride requests, making accurate price prediction based on the clusters, detecting hidden correlations in the data and allowing fast convergence due to the network topology. We provide an empirical evaluation of the application of ADMM network Lasso on real trip record and simulated data, proving their effectiveness since the mean squared error of the algorithm's prediction is minimised on the test rides.
	\end{abstract}
	
	\section{Introduction}
	City councils and private commercial companies with a smart city initiative have recently been aiming for facilitating ride sharing of public and private transportation systems that provide significant environmental benefit in terms of reduced energy consumption and carbon footprint. Ride sharing is a service that arranges shared rides or carpooling on very short notice by better utilization of empty seats in vehicles. of This is especially important during the rush hours when there is a significant surge in demand for public transport leading to long waiting times and higher tariff rates for the commuter. This is when an elevated supply of vehicles aggravates traffic congestion and carbon emissions, while lowering the net income for the drivers when the demand subsequently falls following the rush hour. It is therefore imperative to develop smart ride sharing algorithms to optimize for the best outcome.
	\begin{figure}[h]
		\begin{center}
			\includegraphics[width=0.75\textwidth,clip=true,trim=20 0 50 0]{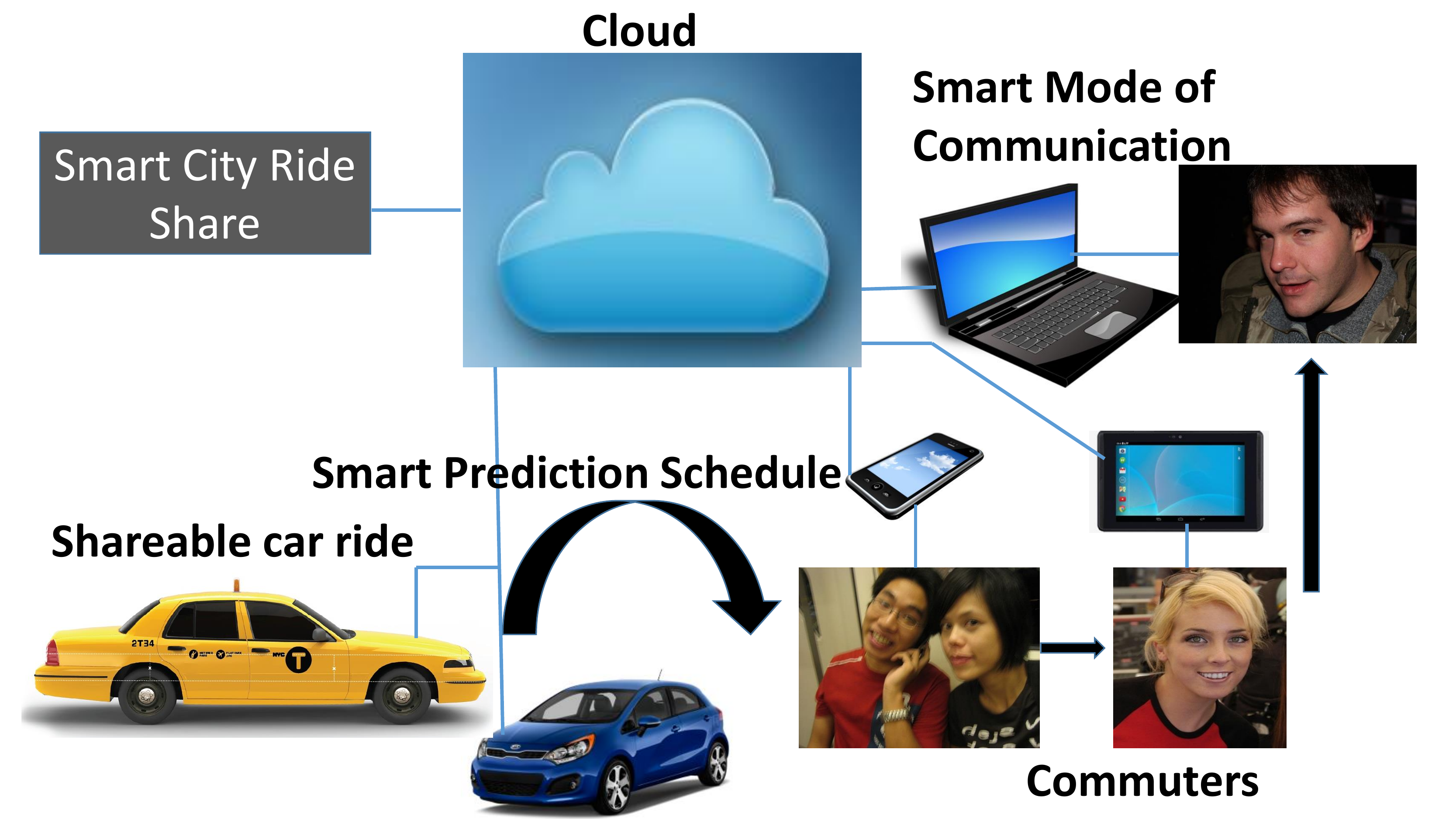}
		\end{center}
		\caption{Smart City Ride Share Prediction System.}
		\label{fig:smartcity}
	\end{figure}
	Traditionally such ride sharing systems are accompanied by a smartphone or tablet based application, with which potential riders can make real time requests. The system then dispatches the ride shared vehicle or taxis for the pick up after a decision making process that usually takes place in the cloud. The vehicles are also equipped with a version of the application that can communicate with the server. Typically, there are two stages to the planning: (i) first vehicles are searched that match the different constraints and criteria of the ride share scenario. For example, search and match the vehicle within $0.2$ mile radius of the ${\tt lat,lon}$ with an available capacity of at least $2$; (ii) the search and match phase is typically followed by the schedule or plan for the pick-up whilst satisfying the minimum increase in distance, costs and maximum profit margin for both the riders and the drivers. The Figure~\ref{fig:smartcity} illustrates such a smart city ride share system with the commuter images taken from the Pascal VOC 2012 challenge~\cite{pascal-voc-2012}.  In reality, most ride requests are generated in real-time almost near the commute time. The requests need to be processed with minimum response time delay whilst addressing the dynamic context such as the current rush hour surge demand among others. In this work we address this quick response and dynamic context of serving shareable rides to new requests by using models learnt on historical data.
	\section{Motivation and Contribution}
	Most ride sharing systems cannot learn models to facilitate dynamic real-time prediction of ride sharing opportunities before the actual scheduling, searching and matching process. Specifically learning a model of interactions and using the same model for new data is novel to the applications in this field. We are motivated by simplifying the search, match and scheduling phases: by bridging the gap in learning from the correlations in the ride or trip data for optimization and clustering followed by efficient predictions before the planning and scheduling stage. Intuitively, this is because if latent groupings are detected in the data, then the search space for the optimization problem is drastically reduced. Additionally, jointly optimizing and clustering can avoid the separation of different phases and save significant delay in response time to a ride request. 
	
	Consider, an example, pick up locations  A {\tt 40.747, -73.893}, B {\tt 40.69, -73.969}, C {\tt 40.82,-73.944}, D {\tt 40.744,-73.912}, A and B are 6.5 miles apart,  A and C are 7.1 miles apart. B and C are 13.5 miles apart (route with tolls), D and C are 8.2 miles apart (route with tolls) and D and B are 5.7 miles apart. Further, the route with tolls have heavy congestion at the time of request C, known from past information. Also, the requests within $0.2$ miles within C have in the past rated the rides shared from around D negatively. Requests from around the region B have found rides shared with region from A expensive. With all this information, a possible clustering is A,C and B,D.  Knowing the implicit clustering before the planning and scheduling phase saves the costs and the delay involved in re-planning, resource allocation, optimizing over all the four rides and serving requests.
	
	Lasso is a statistical machine learning technique that is known for its capabilities for simultaneous variable selection and estimation of the function with added regularization. Automatic variable selection is useful especially when all the variables pertaining to the ride might not carry meaningful information or might not be available. Variants of Lasso can capture correlations between various parameters of A, B, C and D to optimize and cluster them jointly.
	
	To the best of our knowledge this is the first application of a joint optimization, clustering and prediction framework within the ride sharing purview that is fully scalable. 
	\begin{figure}[h]
		\begin{center}
			\includegraphics[width=0.85\textwidth,clip=true,trim=20 70 20 40]{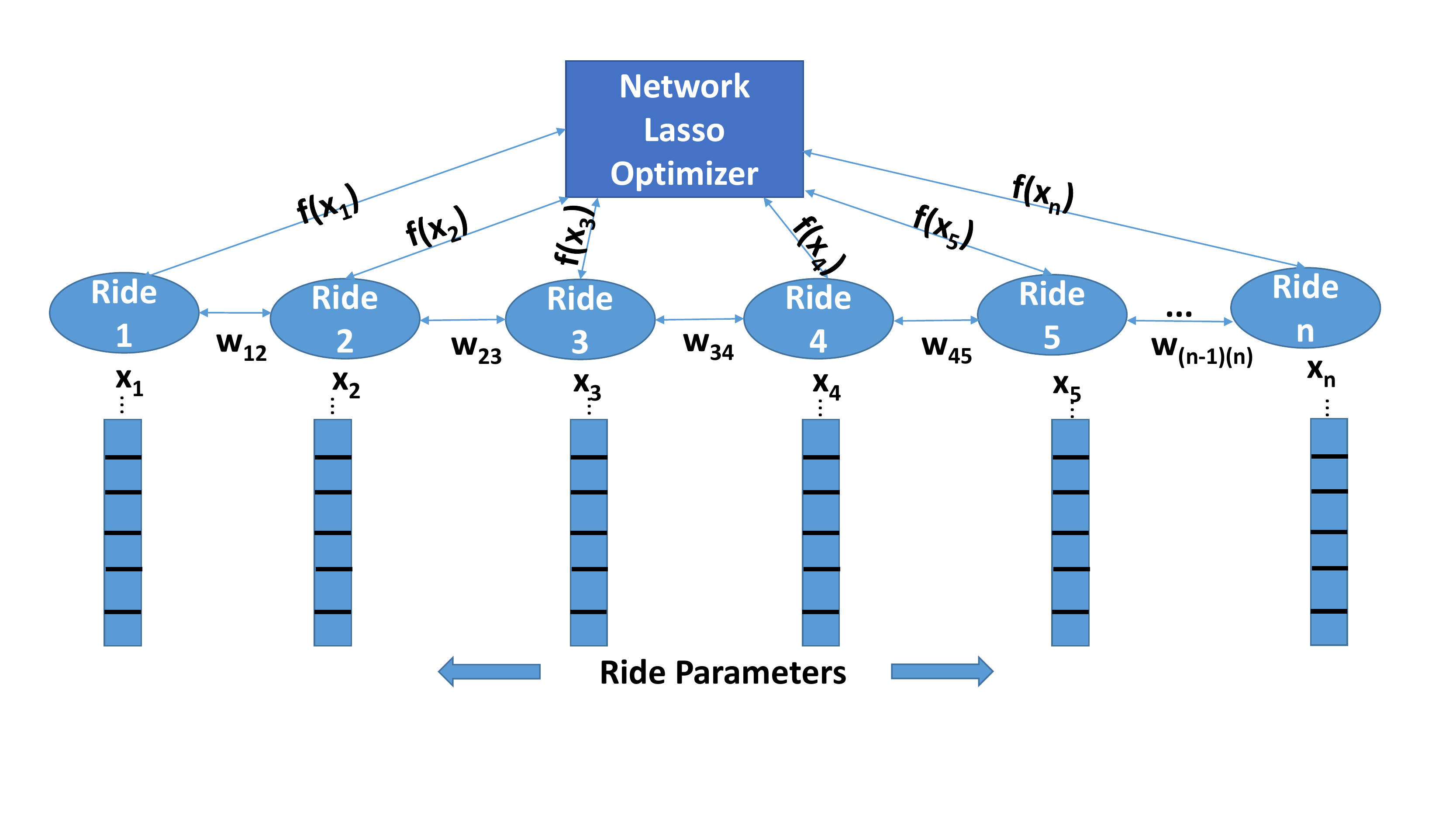}
		\end{center}
		\caption{Network Lasso Ride Share Optimizer.}
		\label{fig:nwlasso}
	\end{figure}
	An emergent property of the system we propose is its ability to  predict the optimal pricing based on the clusters. Such optimal pricing is imperative in case of commercial systems like Uber that apply surge pricing for rush hours when the demand is high. The danger of surge pricing by a multiplicative factor is that it can lead to a reduction of rider interest. If the surge pricing can be predicted beforehand by learning the patterns from the historical data, the supply can be adequately increased or ride share opportunities provided to offset the surge pricing. 
	
	\subsection{Summary of Contributions}
	The main contributions of this paper can be stated as follows:
	\begin{itemize}
		\item 
		We conceive an application of a robust and scalable machine learning enabled large scale ride sharing system that learns a model of correlations from historical trip/ride request data, and uses it to predict ride sharing opportunities on current data. 
		\item
		Applying Network Lasso convex optimization algorithm~\cite{hallac2015network} can jointly optimize model parameters, detect hidden clusters and predict parameter value on test trip/ ride requests thus reducing the search space for any traditional ride sharing system phases to follow if required.
		\item
		The empirical evaluation of the application of Network Lasso algorithm on simulated and real datasets, show the efficient grouping (clustering) of training trip records, deducing test trip record (not yet served) model parameters based on its cluster membership and accurate prediction of its fare pricing.
	\end{itemize}  
	
	\section{Related Work}
	Carpooling systems and recurring ride sharing systems~\cite{carpool1,carpool2} have studied the ride sharing problem, although investigating the daily commute routine only with requests that were preplanned.  In the works that have studied the real-time ride share prediction problem such as~\cite{ma2013t,ma2015real}, the focus of their methods was on the searching and scheduling of the taxis for ride sharing. For example, searching the ride sharing vehicle closest to the pick up points or scheduling the vehicles such that the total distance is minimized. In our work, the main focus is placed on the stage prior to the searching, scheduling and matching riders to drivers. The model learnt during this stage can enable real time prediction at later stages. The research work in the dial-a-ride problems (DARP)~\cite{cordeau2007dial,horn2002fleet} have studied static customer ride requests which are known a-priori. 
	
	Although these methods perform grouping of requests beforehand, requests do not get served real-time. In the work of Zhu et. al.~\cite{7471482}, the authors focus on path planning algorithm for the ride share vehicles with minimized detour. Further, capturing the spatio-temporal underlying features of the rides in our work is at the stage of inferring similarities in rides by means of grouping which is different from the spatio temporal embedding in the work of Ma et. al.~\cite{ma2013t,ma2015real}. They form the topology over rides based on if rides can be shared together. In out work, the topology over rides is used to find  Traditionally, as evident from the works of~\cite{ma2015real}, a Poisson Model is assumed for the distribution over the ride requests. We assume a similar distribution of the ride requests with which we simulate the real time requests for prediction. In the work of Santi et. al.~\cite{santi2013taxi}, a network topology is used for ride share prediction in the similar way as we do. However, our method differs in the optimization that we adapt for ride related data, jointly optimizing on individual objective and neighbouring objectives on the graph. Other literature focuses on efficient scheduling of ride share vehicles (path planning)~\cite{ma2015real,ma2013t}, recommendations for drivers~\cite{recomm1,recomm2}, pricing for commuters~\cite{mobile} and impact of ride sharing~\cite{mobile,ccolak2016understanding} or static grouping of riders~\cite{recomm1,recomm2}. 

	\section{Formulating a Ride Sharing Model}
	Regression analysis is a well known machine learning paradigm in the statistical method of estimation of the relationship among variables. An important criterion of this analysis is establishing how the behaviour of the dependant variable changes with respect to the multiple independent variables.  Supervised regression~\cite{friedman2012fast} is capable of inferring the functional relationship between the output variable and the input variable; the learnt model is then used to predict the output response on new input data. The learning is performed by comparing the quality of the predicted response value from the model that the algorithm comes up with with the true response value by means of a loss function. Progress is made by the algorithm by taking minimizing the loss over all the data that it sees.
	\subsection{Learning the Model by Lasso Regression}
	Lasso based optimization technique is a type of regression analysis that can automatically detect which independent variables are important in influencing the behaviour of the dependant variable. Lasso algorithms~\cite{tibshirani1996regression} can perform feature selection and supervised regression simultaneously.  For example, the variable pertaining to the ride fare might not be available for a particular set of records whilst they might have the drop off location. Lasso can automatically select some of the dependant variables that strongly influences the model from the variables present. In the example, it is capable of selecting the drop off location to interpolate the fare from the distance travelled in the model for those records where it is absent. Additionally, Lasso helps induce the sparse (zeroed variables) representation within the model such that the contribution from some variables can be turned off selectively. For example, in the model, Lasso may completely ignore the contribution of the variable payment-type by setting it to zero as payment-type has no correlation with total fare. This is made possible by the use of a regularization penalty that penalizes for complex models (many non zero variables) over sparse models (many zero variables) such that the model better generalizes to test data. 
	
	Let us consider a social system where $m$ different ride requests that are being orchestrated simultaneously. Let $i$ denote any such ride request that is being optimized within the framework. Let us assume the scenario can be modelled by a function of some linearly independent variables measured over a period of time that describes a ride request or a ride served. Let $y_i = f_{i}(x_i)$ denote the mapping between the variable $x_i$, where $x_i \in \mathbb{R}^d$ encodes the behaviour of the system and $y_i$ is the response or outcome such that $y_i \in \mathbb{R}$. The only assumption on the function $f$ is that it is a convex smooth function. $p$ is number of observations for the pair $(x_i,y_i)$. For simplicity, let $x_i$ encode four independent variables describing a ride such as distance, time-of-day, pick up location and payment-type. 
	The unconstrained Lasso formulation is given by the following equation.
	\begin{equation}\label{eq:lasso}
	\text{minimize}\hspace{0.2cm} \frac{1}{2} ||f(x_{i})-f(\tilde{x_{i}})||_{2}^{2} + \lambda |x_i|_1
	\end{equation}
	
	In Equation~\ref{eq:lasso}, $f(\tilde{x})$ is the predicted response value of the regression model whereas $f(x)$ is the actual value and $\lambda$ is the non-negative regularization parameter. The first term in the quadratic programming formulation measures the loss between the predicted response variable and the true response variable while the second term in the Lasso penalty or L-$1$ norm that ensures the sparsity in the model (minimizing the sum of absolute values of the variables). The output of Lasso is a model that is capable of interpolating the total fare of the ride as a function of the variables of any new ride request.
	\begin{figure}
		\begin{center}
			\subfloat[][]{\includegraphics[width=0.4\textwidth,clip=true,trim=20 160 50 170]{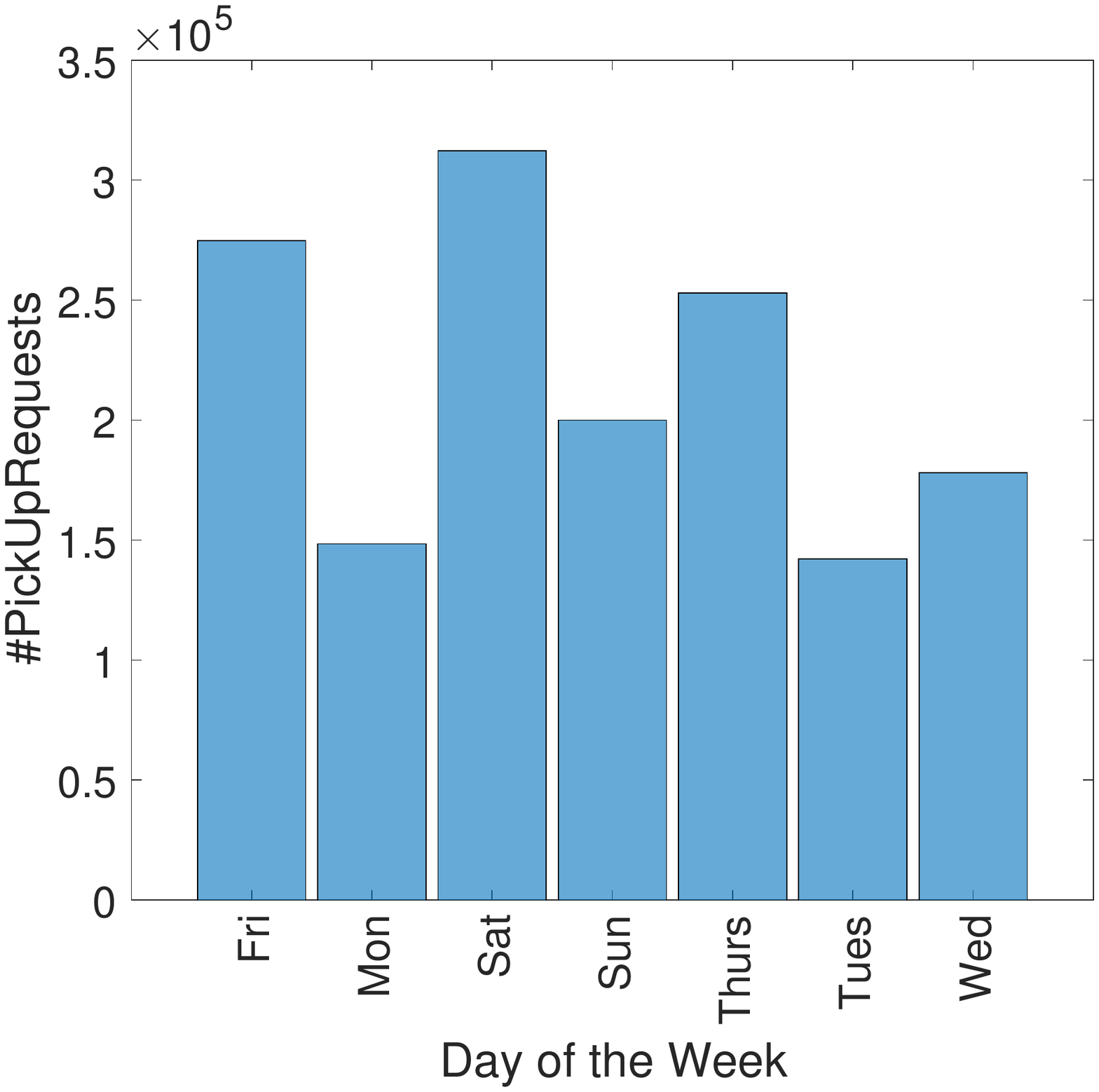}}\vspace{-.3cm}
			\subfloat[][]{\includegraphics[width=0.4\textwidth,clip=true,trim=20 160 50 170]{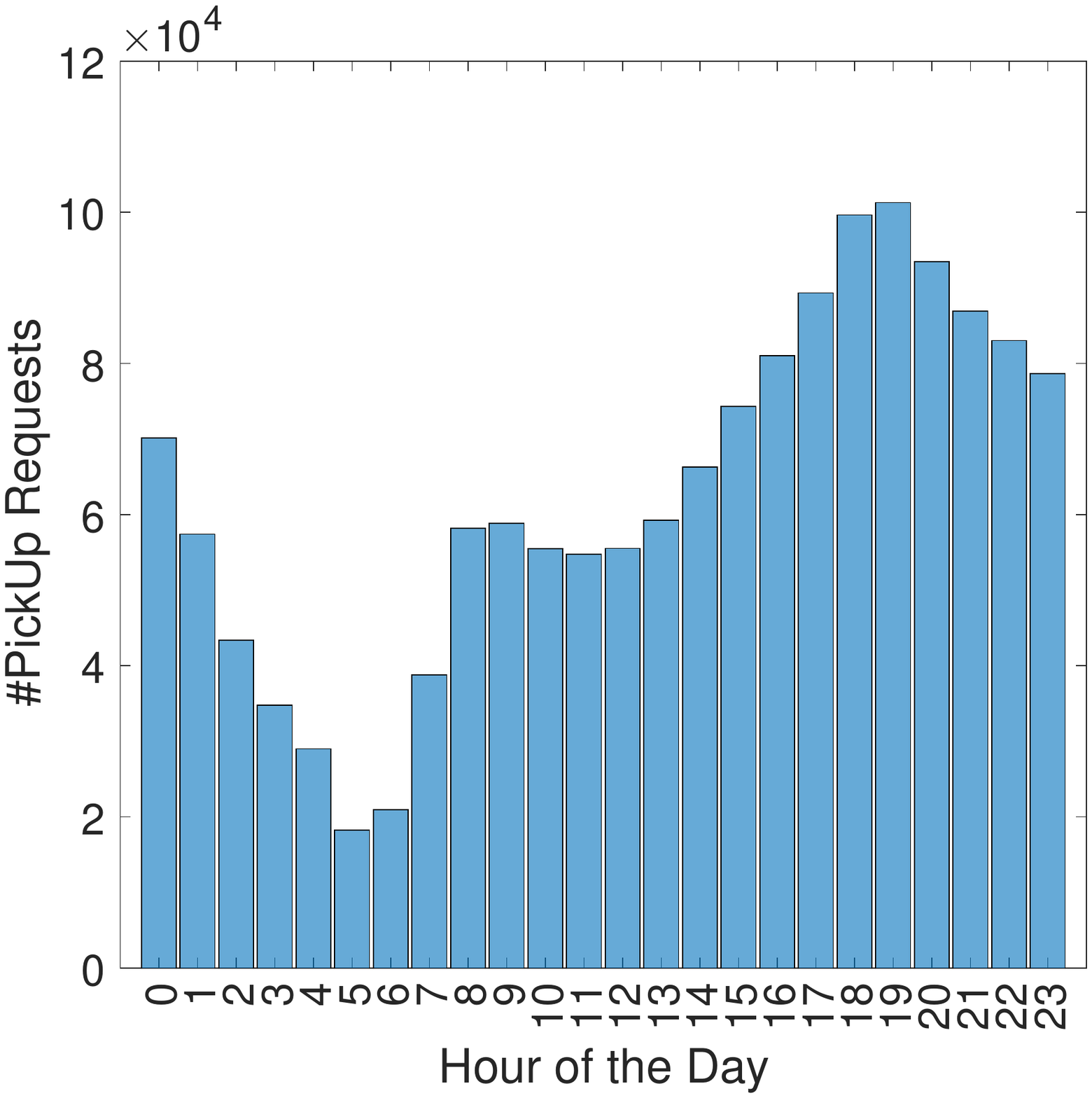}}
		\end{center}
		\caption{Taxi Pick-up request distribution (a) Day of the week (b) Hour of the day.}\label{fig:wkhr}
	\end{figure}
	\subsection{Alternating Direction Method of Multipliers Optimization}
	Typically, for large scale problems with immense datasets where $d << p$ and $d$ is very large, the vanilla Lasso technique is hindered in terms of scaling with the data; that is the optimization becomes extremely time consuming. In our situation, a full fledged social system like ride sharing system may have ride data records in the order of millions of observations of multi-dimensional variables. Essentially, a fully scalable and robust system should scale with the number of observations $p$ and the number of variables $d$. Optimization without exploiting structure in the problem makes the convergence time scale with the cube of the
	problem size~\cite{boyd2004convex}. For such situations, optimization in the primal may be cumbersome and one should resort to optimization techniques in the dual. Dual decomposition ensures that the function can be decomposed and each decomposition can be solved separately. This leads us towards investigating dual decomposition technique for scaling the computation across multiple computing resources. One such technique is Alternating Direction Method of Multipliers (ADMM)~\cite{wahlberg2012admm}. 
	\begin{figure}
		\begin{center}
			\includegraphics[width=0.6\textwidth,clip=true,trim=20 160 50 170]{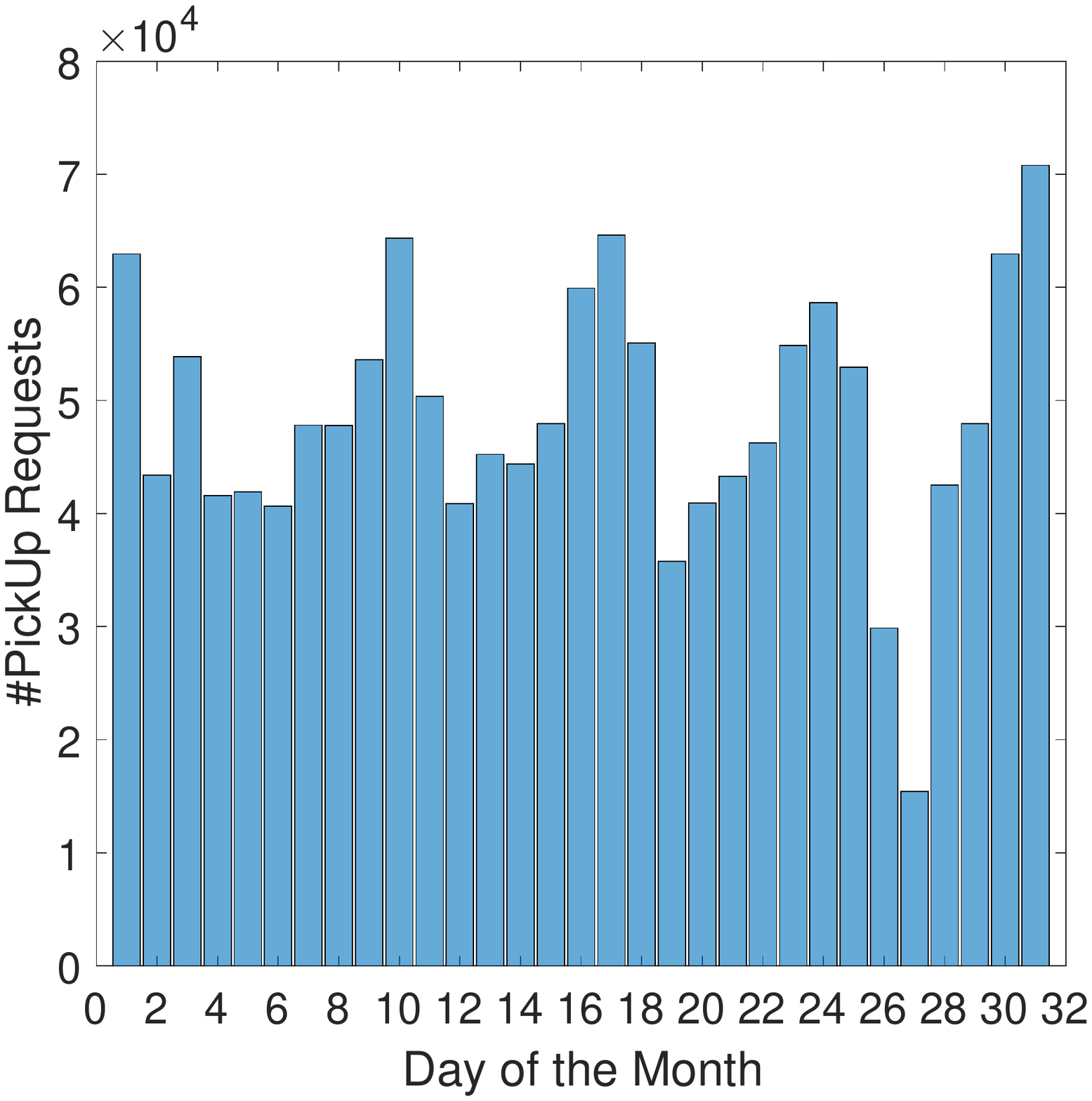}
		\end{center}
		\caption{Taxi Pick-up request distribution over day of the month.}\label{fig:pickday}
	\end{figure}
	This method guarantees the decomposition and robustness with the introduction of new augmenting variables and constraints. In other words, under certain assumptions, additional auxillary variables can be introduced to the optimization problem that enable decomposition of the problem with additional constraints for distributed optimization. This allows scalability and robustness. Let us consider the Lasso model in Equation~\ref{eq:lasso}, within the ADMM model that can be written as:
	\begin{dmath}\label{eq:admmlasso1}
		\text{minimize}\hspace{0.2cm} \frac{1}{2} ||f(x_{i})-f(\tilde{x_{i}})||_{2}^{2} + \lambda |z_i|_1\\
		\text{subject to}\hspace{0.2cm}x_i-z_i = 0.
	\end{dmath}
	where $z_i$ is a copy of $x_i$, is introduced to treat the second term as a decomposable term in the minimization, such that the two terms can be minimized in parallel towards minimizing the global objective. In our example, $z_i$ is the copy of the ride $x_i$. The constraint $x_i-z_i$ is the consistency constraint linking the two copies. From this point onward, when we refer to $f(x)$, we mean for any ride scenario $i$. In our example, this is equivalent to distributing the modelling of the total fare of the ride as a function of the ride's different variables over multiple processing units, such that each ride is being optimized on a different processor in parallel. This allows the model to converge much faster than the non distributed (non ADMM) Lasso.
	As is the case with ADMM, the steps of the optimization can be broken down into a series of update steps derived from the augmented Lagrangian of Equation~\ref{eq:admmlasso1}. We can rewrite $f(x)$ as $f(x)=Ax$, where $A$ is the coefficient matrix, such that the system of linear equations is given by $Ax = b$, where b is the response vector. So Equation~\ref{eq:admmlasso1} can be rewritten as:
	
	\begin{dmath}\label{eq:admmlasso2}
		\text{minimize}\hspace{0.2cm} \frac{1}{2} ||Ax-b||_{2}^{2} + \lambda|z|_1\\
		\text{subject to}\hspace{0.2cm}x-z = 0.
	\end{dmath}
	The augmented Lagrangian of Equation \ref{eq:admmlasso2} is given by:
	\begin{dmath*}\label{eq:auglag2}
		L_{\rho}(x,z,u) = \frac{1}{2} ||Ax-b||_{2}^{2} + \lambda|z|_1 + u^{T}(x-z) + \frac{\rho}{2}||x-z||_{2}^{2}.
	\end{dmath*}
	where $L$ is the Lagrangian, $u$ is the Lagrange multiplier, and $\rho$ is the cost for violating the consistency constraint.
	Minimizing with respect to $x$ and $z$ separately, and $u$ jointly, leads to the iterates. The iterates can be updated in a distributed way yielding the scalability for very large problems. The main advantage in using ADMM based Lasso besides robustness and scalability is in its guaranteed global convergence.
	\begin{align*}
	&\begin{aligned}
	x_{k+1} 
	= (A^{T}A + \rho I)^{-1}(A^{T}b+\rho z_k - u_k)
	\end{aligned}\\
	&\begin{aligned}
	z_{k+1} = S_{\frac{\lambda}{\rho}}(x_{k+1}+u_{k}/\rho)
	\end{aligned}\\
	&\begin{aligned}
	u_{k+1} = u_k + \rho(x_{x+1}-z_{k+1}).\\
	\end{aligned}
	\end{align*}
	The $x$ update can be derived by minimizing with respect to $x$. We know $||u||_{2}^{2} = u^{T}u$, $(Ax - b )^{T} = x^{T}A^{T} - b^{T}$, $u^{T}v = v^{T}u$
	\begin{align*}
	L(x) = \frac{1}{2}(Ax -b )^{T}(Ax-b) + \lambda|z| + u^{T}x -  u^{T}z + \frac{\rho}{2}(x-z)^{T}(x-z)\\
	= \frac{1}{2}(x^{T}A^{T}Ax - x^{T}A^{T}b - b^{T}Ax + b^{T}b) + \lambda|z| + u^{T}x -  u^{T}z \\+ \frac{\rho}{2}(x^Tx - x^Tz - z^Tx + z^Tz)\\
	= \frac{1}{2}(x^{T}A^{T}Ax - 2x^{T}A^{T}b + b^{T}b) + \lambda|z| + x^{T}u - u^{T}z \\ + \frac{\rho}{2}(x^Tx - 2x^Tz + z^Tz)\\
	= \frac{1}{2}x^{T}A^{T}Ax - x^{T}A^{T}b + \frac{1}{2}b^{T}b + \lambda|z| + x^{T}u - u^{T}z \\+ \frac{\rho}{2}x^Tx - \rho x^Tz +  \frac{\rho}{2}z^Tz\\
	\end{align*}
	We know from the quadratic term $\frac{\partial}{\partial_x}x^TMX = 2Mx$, and the linear term $\frac{\partial}{\partial_x}x^{T}a = a$ where $a$ is a vector. So we have:
	\begin{align*}
	\frac{\partial L_{\rho}(x,z,u)}{\partial x} = A^{T}Ax + \rho{x} - A^Tb + u -\rho z\\
	\implies (A^{T}A + \rho I)x - (A^Tb + \rho z - u)  = 0\\
	x =  (A^{T}A + \rho I)^{-1}(A^{T}b + \rho z - u)
	\end{align*}
	Hence, $x$ can be updated at time $k+1$ with values of iterates $z$ and $u$ from time $k$. $A^TA + \rho I$ is positive definite and hence invertible.

	Similarly, the $u$ update is also obtained by minimizing with respect to $u_k$ for $x_{k+1}$ and $z_{k}$. The update for $z_{k+1}$ is obtained from a soft shrinkage solution.
	In the simulation experiments that we discuss in~\ref{sec:lassosyn}, we show how the Lasso optimization model is capable of estimating the response variable with respect to the input  variables. The Lasso is used mainly in an optimization problem for inferring the model of the variables.
	\begin{figure}
		\begin{center}
			\includegraphics[width=0.6\textwidth,clip=true,trim=0 0 0 0]{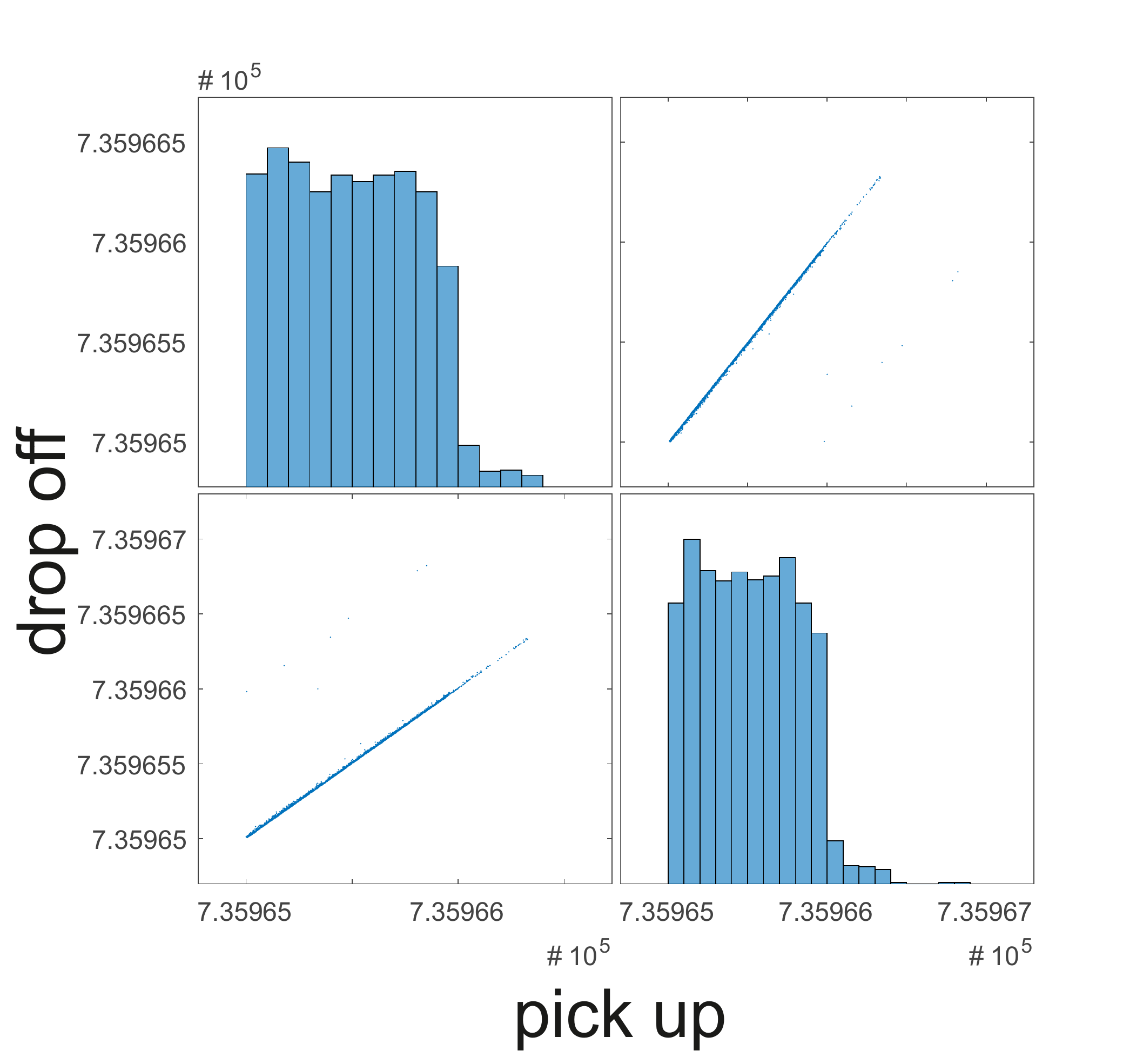}
		\end{center}
		\caption{Correlations between the pick up and drop off times of trip parameters with pick up times on the x axis and drop off times on the y axis.}
		\label{fig:timedur}
	\end{figure}
	As we shall see in section~\ref{sec:lassosyn}, ADMM Lasso is capable of modelling the utility of rides as function of different synthesized ride parameters using distributed optimization.
	\subsection{Network Lasso ADMM Optimization}
	The Network Lasso~\cite{hallac2015network} algorithm is a generalized version of the Lasso to a network setting that enables simultaneous optimization and clustering of the observations.  Network Lasso~\cite{hallac2015network} extends the power of Lasso algorithms through structured learning over a graph topology. The topological structure allows for joint optimization. In our example, this means groups of rides gets automatically clustered and optimized to have the same models of the total fare as a function of their ride parameters. Not only the grouping of rides  is performed based on similar rides being grouped together, the optimization problem also computes the fare model across such groups in a consensus.
	
	Let a graph be described by $\mathcal{G} = (\mathcal{V,E})$, where $\mathcal{V}$ is the set of vertices and $\mathcal{E}$ is a set pf edges connecting neighbouring vertices. The graph or network as shown in Figure~\ref{fig:nwlasso} encodes the input data such that each data point is represented as a vertex. In our case, each vertex represents a ride trip record or a ride request. The similarity between the trip records or requests is encoded as an edge. The objective that Network Lasso tries to solve is expressed in the Equation below. The variables are $x_i,...,x_m \in \mathbb{R}^p$, where $m = |\mathcal{V}|$ is the number of trip records or ride requests, and $p$ is the number of features of the ride, with a total of $mp$ variables for optimization~\cite{hallac2015network}.
	\begin{equation}\label{eq:nwlasso}
	\text{minimize}\sum_{i \in \mathcal{V}}f_i(x_i)+\lambda\sum_{(j,k)\in \mathcal{E}}w_{jk}||x_j-x_k||_2.
	\end{equation}
	Similar to Hallac et. al.~\cite{hallac2015network}, the function $f_i$ at node $i$ is the local objective function for the data point $i$ whereas the $g_{jk} = \lambda w_{jk}||x_j-x_k||_2$ is the global objective function associated with each edge with $\lambda \geq 0$ and the weight over the edge (a measure of similarity) $w_{jk} \geq 0$. The edge objective function penalizes differences between the variables of the adjacent nodes thus inducing similar behaviour; leading to groups of nodes or clusters that behave similarly; the solution to the optimization problem is the same across all nodes $x_i$ in a cluster. In other words, each cluster has the same model (functional mapping between the ouput and input variables). In our example this would imply similar ride share records get grouped into clusters and hence similar plans and schedules can be allocated to these clusters. The only assumption is on the convexity of the function $f_i$. 
	
	It is important to note the role of $\lambda$, which is a regularization parameter to control the optimization process. Based on the value of $\lambda$, the optimization process trades off optimzing for the node variables and edge variables. The range of values of $\lambda$ determine the level of optimization. For smaller values of $\lambda$, optimization is performed at the node level while for larger values, the edge optimization comes into play inducing the adjacent nodes to have similar model.  The edge cost is the sum of norms on how different the adjacent ride records are from each other and the penalty that needs to be paid within the model for large differences. In other words, the edge objectives encourage nodes to be in consensus (have a similar model). In our application, we use the regularization such that the edge penalties are active. The vanilla Lasso technique discussed above, essentially maps to the scenario where $\lambda = 0$, when the individual nodes are optimized independently without the edge optimization. With the Network Lasso formulation~\ref{eq:nwlasso}, we not only achieve a robust, scalable and distributed optimization algorithm,  but we are guaranteed to obtain global convergence. The edge objective function is the  $g_{jk} = \lambda w_{jk}||x_j-x_k||_2$ adds the ``network'' aspect to the vanilla Lasso optimization, by inducing a relationship between individual node variables. In fact, in our example, this edge objective minimization allows for clustering of rides based on their optimization models.
	In the following section we evaluate Lasso, ADMM and Network Lasso techniques with synthesized and real dataset of ride observations illustrating the accuracy of the optimization while presenting ride sharing opportunities.
	\section{Model Validation and Experimental Evaluation}
	In this section, we discuss the experiments conducted on synthetic and real world datasets to validate the efficiency of the techniques that we propose in the previous section. The section begins with a description of the synthetic experiment we design in order to evaluate the modelling, feature selection and prediction accuracy of the application of the vanilla ADMM Lasso technique to an unknown linear model of variables encoding a ride. Following this discussion, we explain the real dataset experiments, where the open trip record dataset of the green taxis from the New York Taxi and Limousine Commission~\cite{nyc-2015} is used.  We apply Network Lasso on this dataset to enhance the capabilities of vanilla ADMM Lasso in modelling, while being in consensus with the models of the neighbourhood trips.  
	
	\subsection{Synthetic Dataset and Experiment}\label{sec:lassosyn}
	The synthetic dataset is constructed by exploring the linear relationship between the multidimensional variables of the underlying ride sharing model that we assume. For n simplified linear model for a ride request $i$,  
	\begin{align}\label{eq:simadmm}
	f_i(x_i) = a_i.\tt{ratings} + b_i. \tt{preferences}+ c_i.{\tt pickuptime} \\ \nonumber + d_i.{\tt pickuploc} - e_i.{\tt cost}
	\end{align}
	we assume variables such as ${\tt ratings} \in [1,10] \subset {\mathbb Z}$, ${\tt preference} \in [1,10] \subset {\mathbb Z} $, ${\tt pickuptime} \in \{0,1\}$, ${\tt pickuploc} \in [0,30] \subset \mathbb{R}$ and ${\tt cost}$ uniformly distributed $\in [0,1] \in \mathbb{R}$. The variable {\tt ratings} encodes the past feedback of the shared ride experience rating in the past. Variable {\tt preferences} are the choices of the commuter for example sharing with more than 1 or sharing with 1,  {\tt pickuptime} and {\tt pickuploc} are the requested time and location of pickup respectively while {\tt cost} is the expenses related to the ride. The value $f_i(x_i)$ encodes the utility value of the ride as a function of all the variables.
	These variables or regressors are related in terms of parameters of the ride given by $[a_i, b_i, c_i, d_i]$, which is not available to the algorithm. It should be able to deduce the latent model based on the given final fare values $f_i(x_i)$ and the regressors. The application of the ADMM Lasso algorithm enables learning the relationship between the ride variables by modelling them efficiently. The model is improved over many such ride request data records to minimize the error between the internal algorithm model and the true model in hindsight. To this end we apply the Lasso regression technique discussed before in Equation~\ref{eq:lasso}. 
	\begin{figure}
		\begin{center}
			\subfloat[]{\includegraphics[width=0.28\textwidth,clip=true,trim=5 0 5 0]{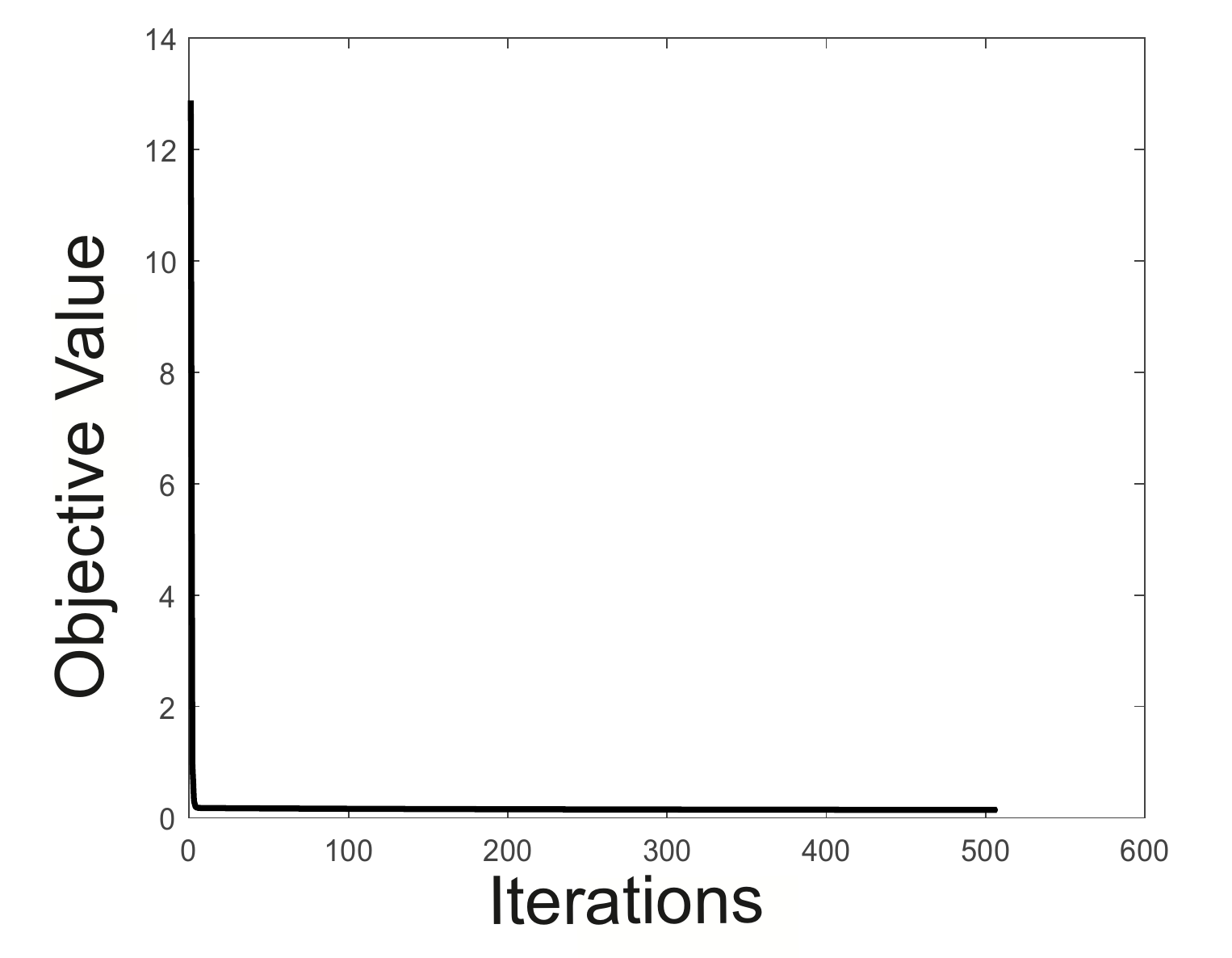}}
			\subfloat[]{\includegraphics[width=0.28\textwidth,clip=true,trim=5 0 5 0]{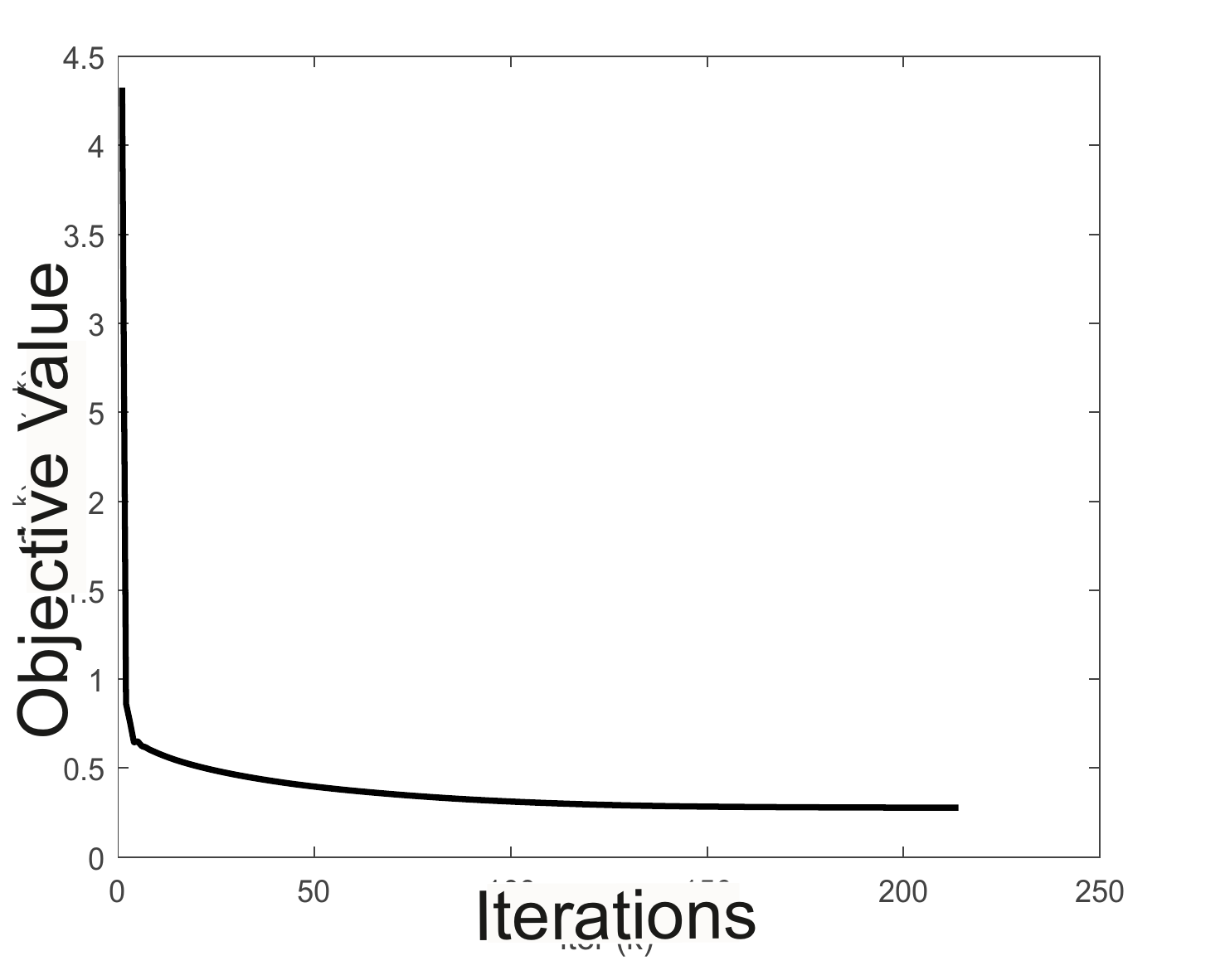}}
			\subfloat[]{\includegraphics[width=0.28\textwidth,clip=true,trim=5 0 5 0]{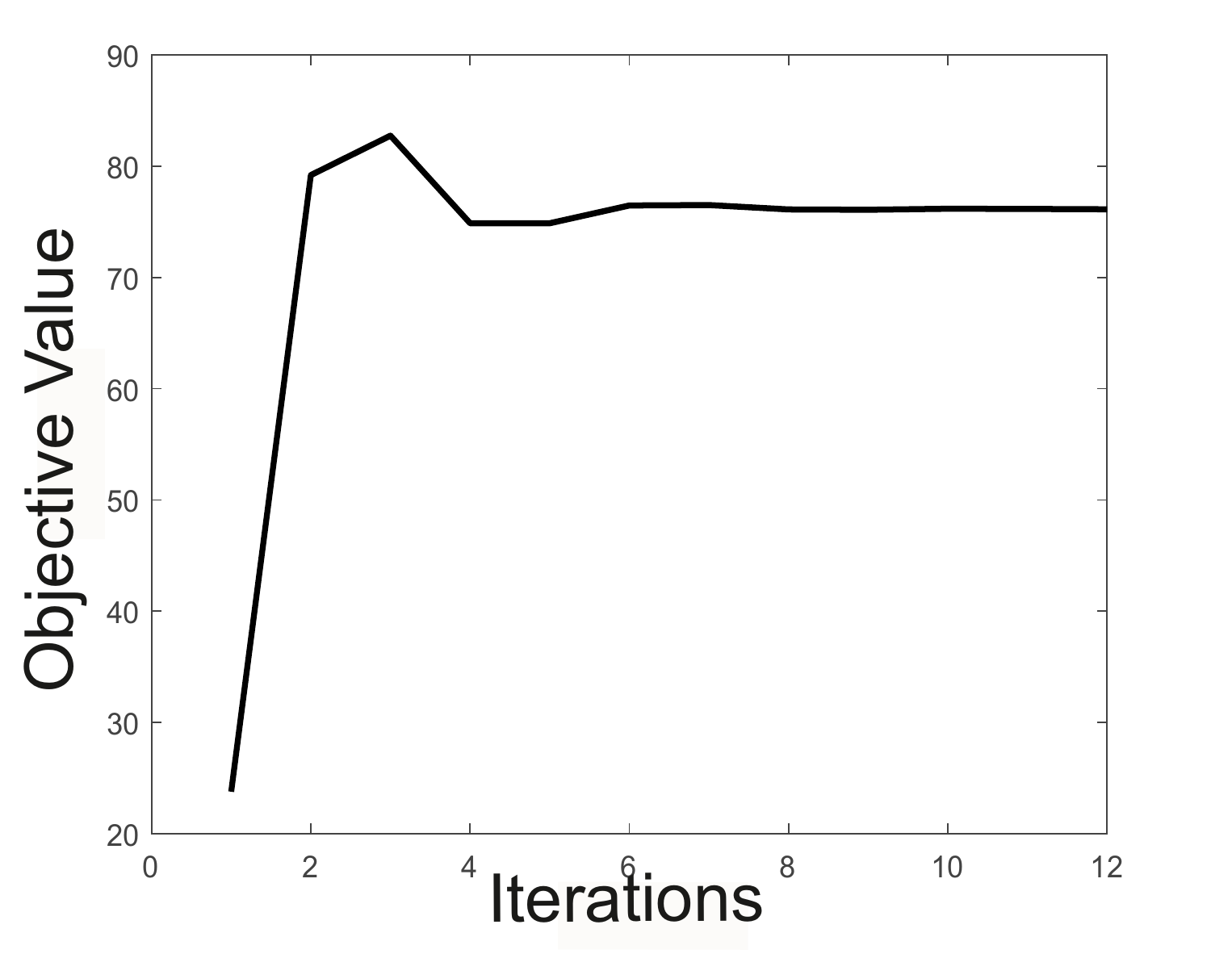}}
		\end{center}
		\caption{Convergence of Lasso ADMM over Iterations for Varying $\lambda$ (a)$\lambda = 0.0001$ with very early convergence but complex model (b)$\lambda = 0.001$ convergence at 50 iterations with simpler model (c)$\lambda = 0.01$ with no convergence at the minimum objective value.}\label{SimResultsObj}
	\end{figure}
	
	For robustness we randomly sample $1500$ examples with $5000$ features. The coefficient matrix $A$ constituting the ride variables is generated as a Gaussian distributed sparse matrix with a known sparsity density, we use a sparsity density of $0.02$. The output variable $b$ corresponding to the input variables is computed as a linear combination between the coefficient matrix and a random sparse vector and some Gaussian noise. 	We vary the value of the Lasso parameter $\lambda$ in a range of $[0.1, 0.001, 0.0001]$, to evaluate the influence on the smoothness of the optimization. The values of the ADMM parameters $\rho$ is fixed to $1.2$ and that of $\alpha$ is fixed to $1.8$. For all our simulation experiments, we adapt the ADMM Lasso code~\cite{lassocode} for our data, the code for which is written by us. The experiments were carried out on a Windows Desktop PC with 16GB RAM and i7 processor using Matlab.
	
	The results of our simulation experiments are shown in Figure~\ref{SimResultsObj}.  Higher values of $\lambda$ induce more sparsity, by penalizing complex models thus allowing simpler model where most of the variables are zeroes. This is desirable for generalization and high accuracy on any test time (new) data. Lower values of the penalty parameter allow for denser solutions with more non-zeroes. A trade-off is often desired to have simpler models that do not over-fit the training data and that generalizes to test data. In the Figure~\ref{SimResultsObj}, the value of minimum of the loss function in Equation~\ref{eq:lasso} (vertical axis) on all the plots across time (horizontal axis); the lower the difference between the predicted fare value and the ground truth fare value, the better. In (a), the solution converges quickly for $\lambda = 0.0001$. However this model is complex and can overfit the data on the test data. The total number of non zero variables are $3062$.  In (b) with $\lambda=0.001$, the solution takes longer to converge than (a) with higher prediction error than (a). The number of non-zeroes is 1673 which is about 30 percent of the variables out of 5000. This shows how Lasso is capable in capturing the 30 percent most important ride variables that contribute to the model. In (c), for higher $\lambda=0.01$, the solution is very sparse, however the error increases as the model in unable to fit the data, with number of non-zeroes being only 636. It is important to note that although Lasso by itself can induce generalized solutions, the use of ADMM approach for large datasets is desirable for faster convergence as shown here where the convergence happens within 50 iterations in~\ref{SimResultsObj} (b) .
	
	\subsection{New York Taxi Data Experiment}
	\subsubsection{Dataset}
	The real-world dataset with attributes as shown in Table~\ref{table:nyc} constitutes about one billion records of various taxi trips recorded over the entirety of 2015~\cite{nyc-2015} in the city of New York. Here, we only use the green taxi trip records for the month of January, 2015 to conduct our experiments. Each record in the dataset pertains to a ride that was served by a green taxi. The various attributes of the ride are defined by variables such as pick up time, drop off time, pick up location, drop off location, base fare, tips, tax, passenger count, trip distance, trip type among $20$ other variables. It is important to note that the green cabs do not serve on the Manhattan area as we will see later on the plots overlaid on the maps. 
	
	We perform an initial visualization of the dataset for any obvious data pattern. 
	\begin{figure*}[t]
		\begin{center}
			\includegraphics[width=1.0\textwidth,clip=true,trim=0 0 0 0]{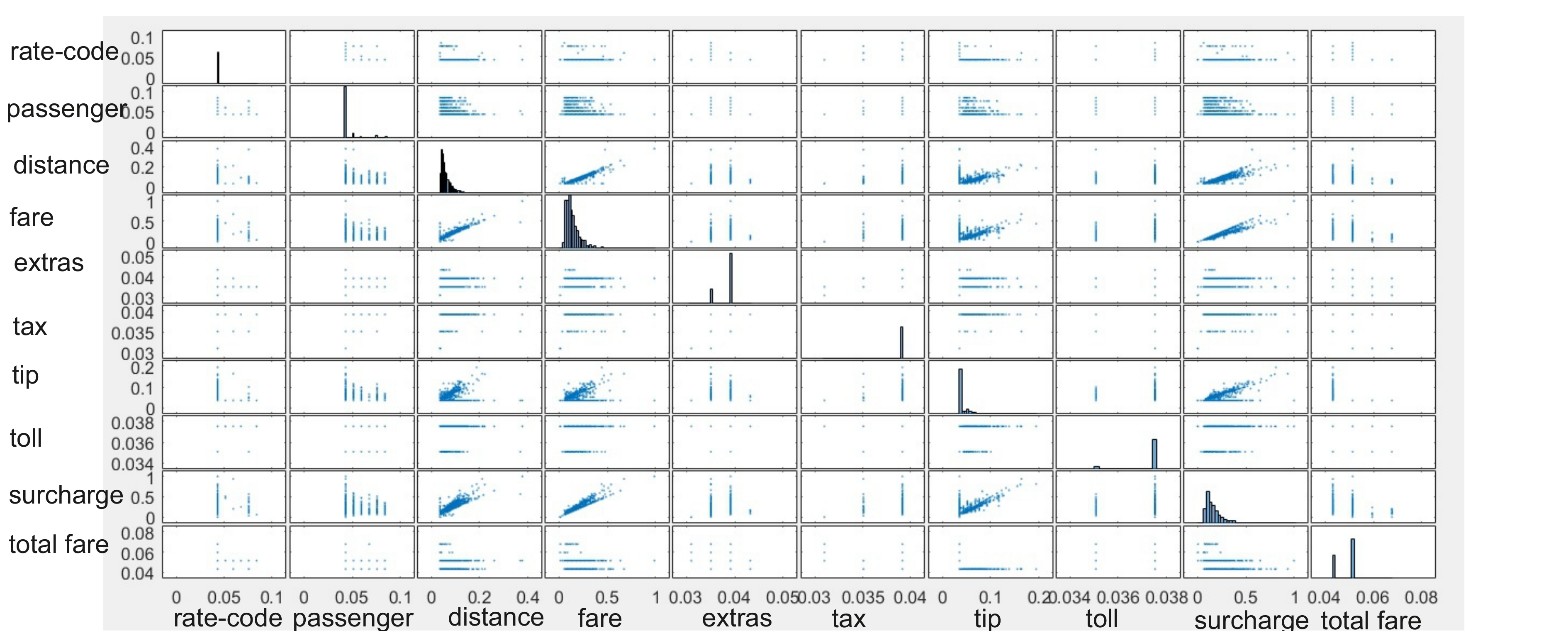}
		\end{center}
		\caption{Correlation across different trip record optimization parameters.}\label{fig:corrfea}
	\end{figure*}
	In Figure~\ref{fig:wkhr} shows the distribution of the users asking for a taxi ride. The requests are plotted for each day of the week and every hour of the day. As expected, we observe from the distribution over days of the week in (a) that shows the weekend and Fridays having an increased demand of taxis. On the plot of the distribution over hours of the day in (b) we observe that there is sharp increase in demand during the rush hour as well as there is a surge in taxi demand during the evening from rush hour through the evening. We observe that most demand is in the early morning hours, when it is difficult to take public transport, continuing into the morning rush hour. The second surge is in the evening rush hour that peaks at around 19:00 hours. The plot over the week is a random week in the month of January, while the plot over the day is a random day of the month of January.
	\begin{table}
		\caption[NYC Dataset]{NYC Taxi Dataset Attributes }\label{table:nyc}
		\begin{center}
			\scalebox{1.0}{
				\begin{tabular}{| l | l | l | l |}
					\hline
					vendorId & pickuptime & dropofftime & storeflag \\
					\hline
					ratecode & pickuplong & pickuplat & dropofflong \\
					\hline
					dropofflat & passengercnt & tripdistance & fareamt \\
					\hline
					extra & mtatax & tipamt & tollamt \\
					\hline
					ehailfee & surcharge & totalamt & paytype\\
					\hline
					triptype &  & & \\
					\hline
				\end{tabular}}
			\end{center}
		\end{table}
		The Figure~\ref{fig:pickday}, we plot the distribution of ride requests throughout the first month in January, 2015. The $27$th day of the month is an exception as there were severe travel restrictions on that day due to heavy snowfall. 
		\begin{figure}
			\begin{center}
				\includegraphics[width=0.6\textwidth,clip=true,trim=20 210 90 160]{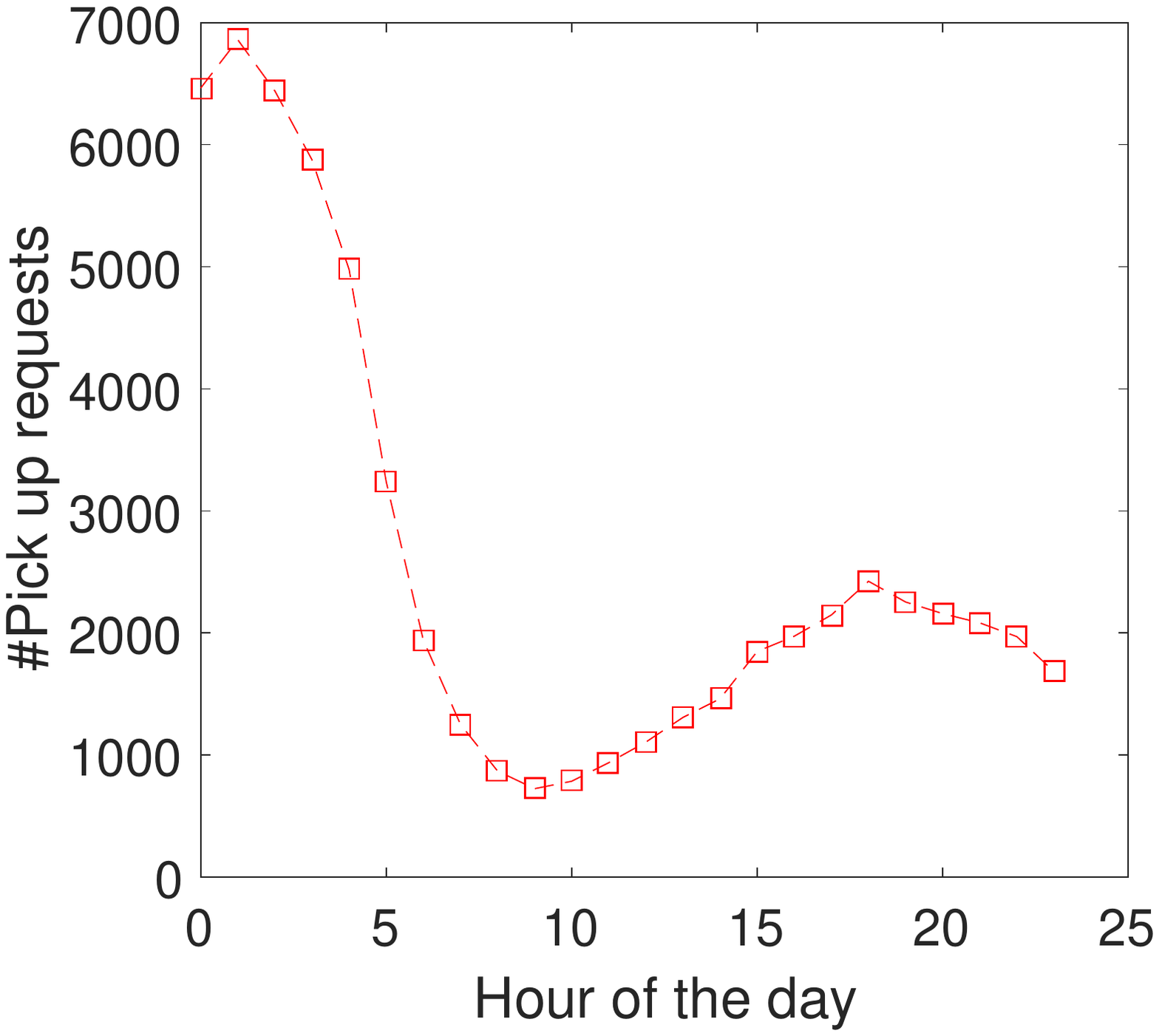}
			\end{center}
			\caption{Frequency distribution of pickup requests over the hour.}
			\label{fig:hourreq}
		\end{figure}
		
		Figure~\ref{fig:timedur} shows the correlations between the pick up time and drop off time of over $2000$ ride requests randomly sampled over days $1$ and $2$ that is used for modelling. As expected the pick up and drop off times are positively correlated. In Figure~\ref{fig:corrfea}, the correlation between different feature variables are shown. As expected the variables fare amount and trip distance are positively correlated with each other. The surcharge is also positively correlated with the tripdistance and the fare amount. The tip in turn is positively correlated with the trip distance and the fare amount.  Figure~\ref{fig:hourreq}, shows the pick up requests generated on the first day of the month of January 2015 per hour. What is interesting is that the distribution of the pick ups behave like a Poisson distribution based on the nature of the curve. In practice, Poisson distribution is often used to model pick up requests. Since we have such a distribution available from the data, it is practical that we use this realistic distribution to sample our test data to evaluate the prediction of the algorithms.
		\begin{figure*}[t]
			\begin{center}
				\includegraphics[width=0.67\textwidth,clip=true,trim=60 210 85 200]{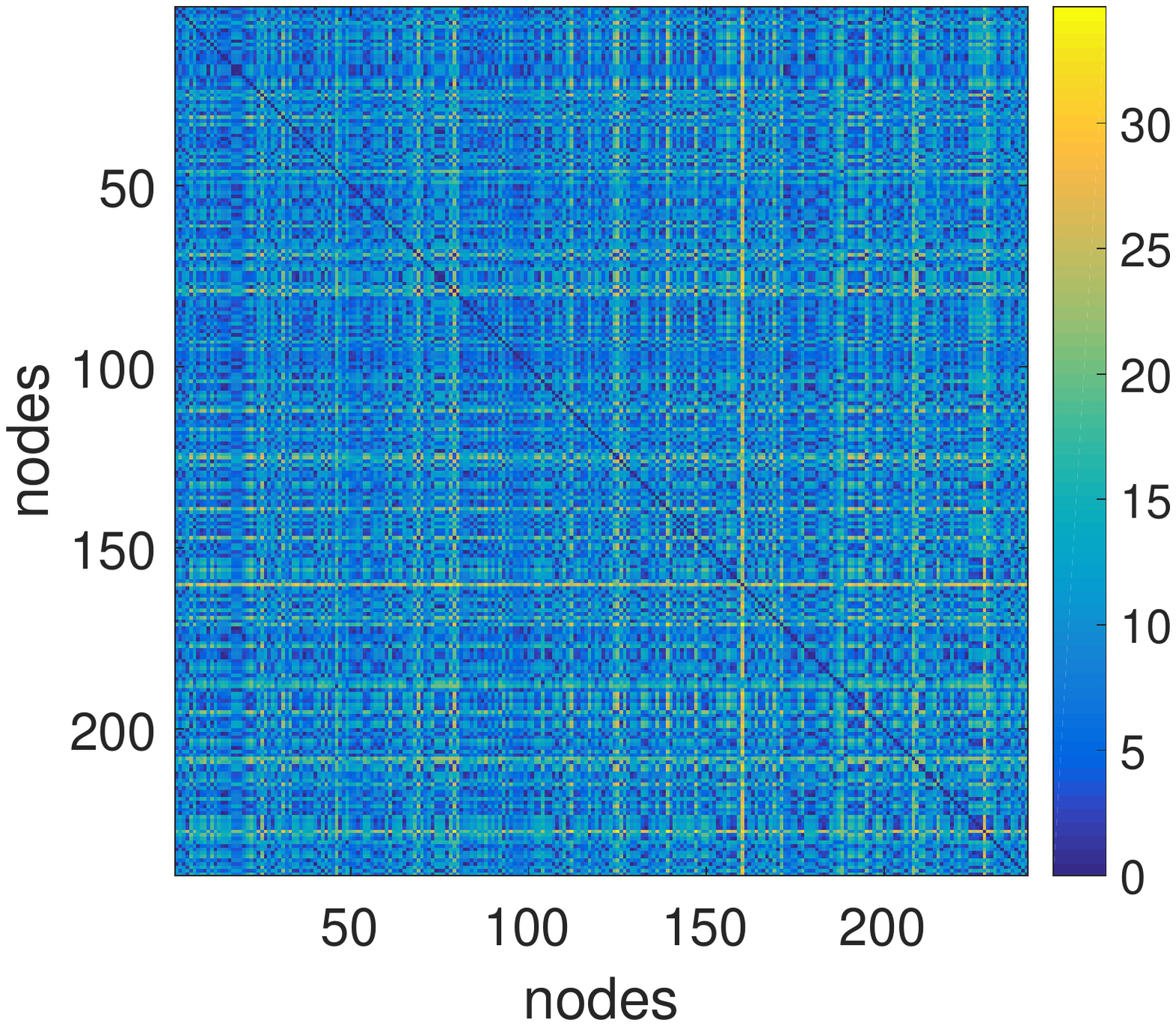}
			\end{center}
			\caption{Heat map of the network derived from the random subset of trips.}\label{fig:heatmap}
		\end{figure*}
		
		In the Figure~\ref{fig:heatmap}, we observe the heatmap result from the network obtained from the dataset. The network is a relatively dense network with distances varying between 0 kilometre (distance with itself) indicated in dark blue to 30 kilometres indicated in yellow. The map indicates ride parameters that are closer to each other in similar colour.
		
\begin{figure*}[t]
			\begin{center}
				\includegraphics[width=0.67\textwidth,clip=true,trim=0 120 0 80]{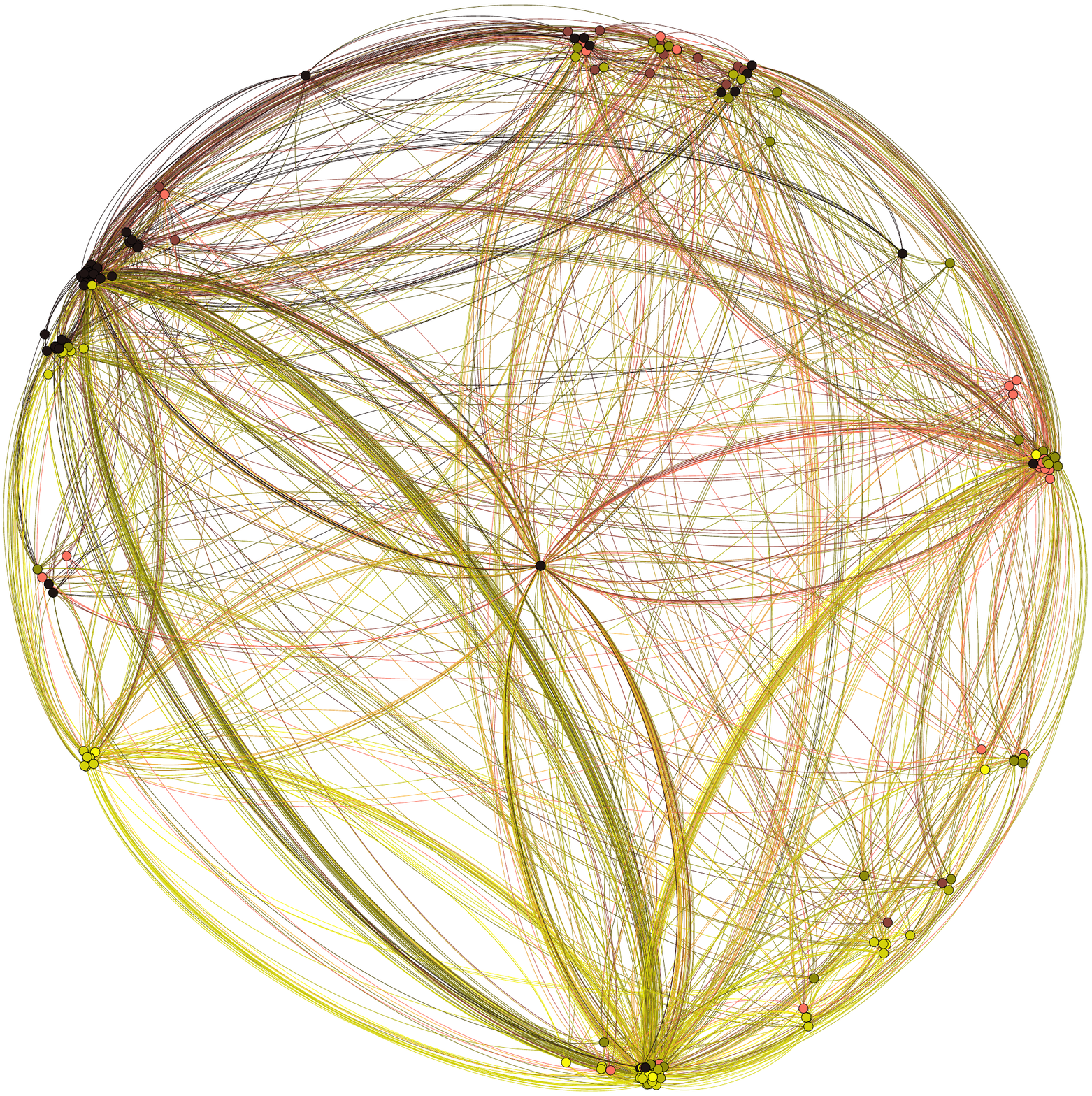}
			\end{center}
			\caption{Spatial clusters without optimization. Network rendered using Gephi software. Edge colours same as source cluster membership colour.}\label{fig:smartsocgephi16}
		\end{figure*}

		In Figure~\ref{fig:smartsocgephi16}, we show the network as generated from the data. The network is based on the spatial information encoded in the data. Basic modularity based on this spatial information shows how the network is formed of dense clusters. Each cluster is indicated by an individual colour and $8$ clusters are formed each indicated with a different colour. It is interesting to note that since these are only spatial clusters, optimizing based solely on these clusters would not factor in consensus in the models. For example, two spatially distant rides, may have the same underlying model of optimization with parameters behaving similarly. 
		%
		\subsubsection{Network Lasso Experiments}
		We apply the network lasso technique to data sampled randomly from the taxi trips dataset described in the previous section. Our training set comprises random subsets sampled from different times of the first and second days of the month. The Network Lasso algorithm~\cite{hallac2015network} learns the model on the training set with a known output response variable. Once the model is learnt, the prediction of the model is evaluated on a test set. The test set is again randomly sampled; the test set does not include any training data. The optimized data attributes are all the attributes other the spatio temporal data. The spatio temporal data is fed as the network information to the algorithm; each trip record is a node in the network. The algorithm optimizes for the total fare value at each node while ensuring consensus among neighbourhood data. The result is a grouping of the network into clusters with similar models. The advantage is these clusters can be used to predict on any new data. Here, for every instance in the test set, the error between the predicted total fare value and the true total fare value is calculated, and the mean squared error is reported for different values of $\lambda$. Varying $\lambda$ tunes how much of consensus is desired between the node models. We adapt the code provided by Hallac et al.~\cite{nwlassocode}. All the experiments are run on a Linux desktop PC with 12 GB RAM and i5 processor using Python.
		
\begin{figure*}[t]
			\begin{center}
				\subfloat[][]{\includegraphics[width=0.45\textwidth,clip=true,trim=20 180 12 220]{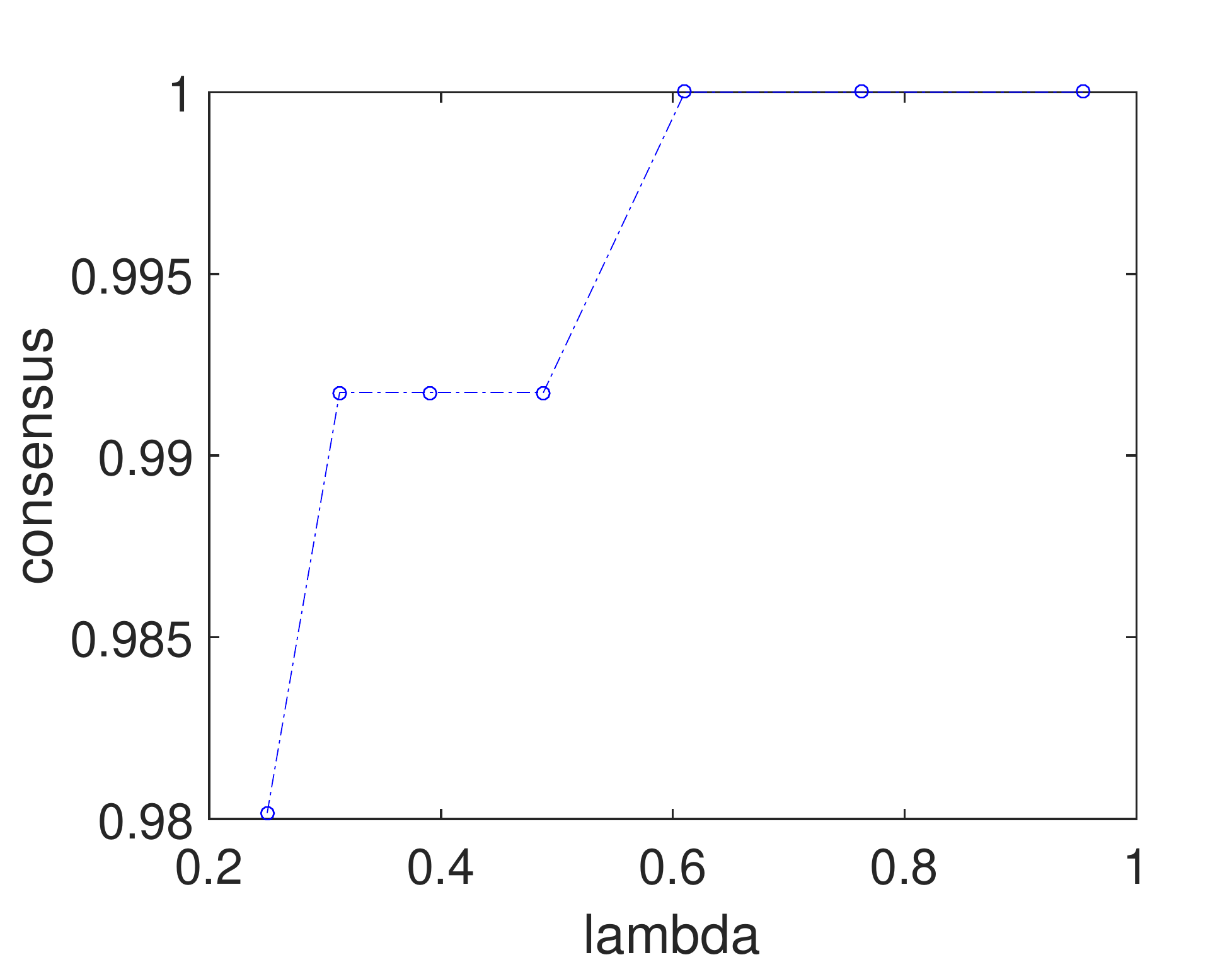}}
				\subfloat[][]{\includegraphics[width=0.45\textwidth,clip=true,trim=20 180 12 220]{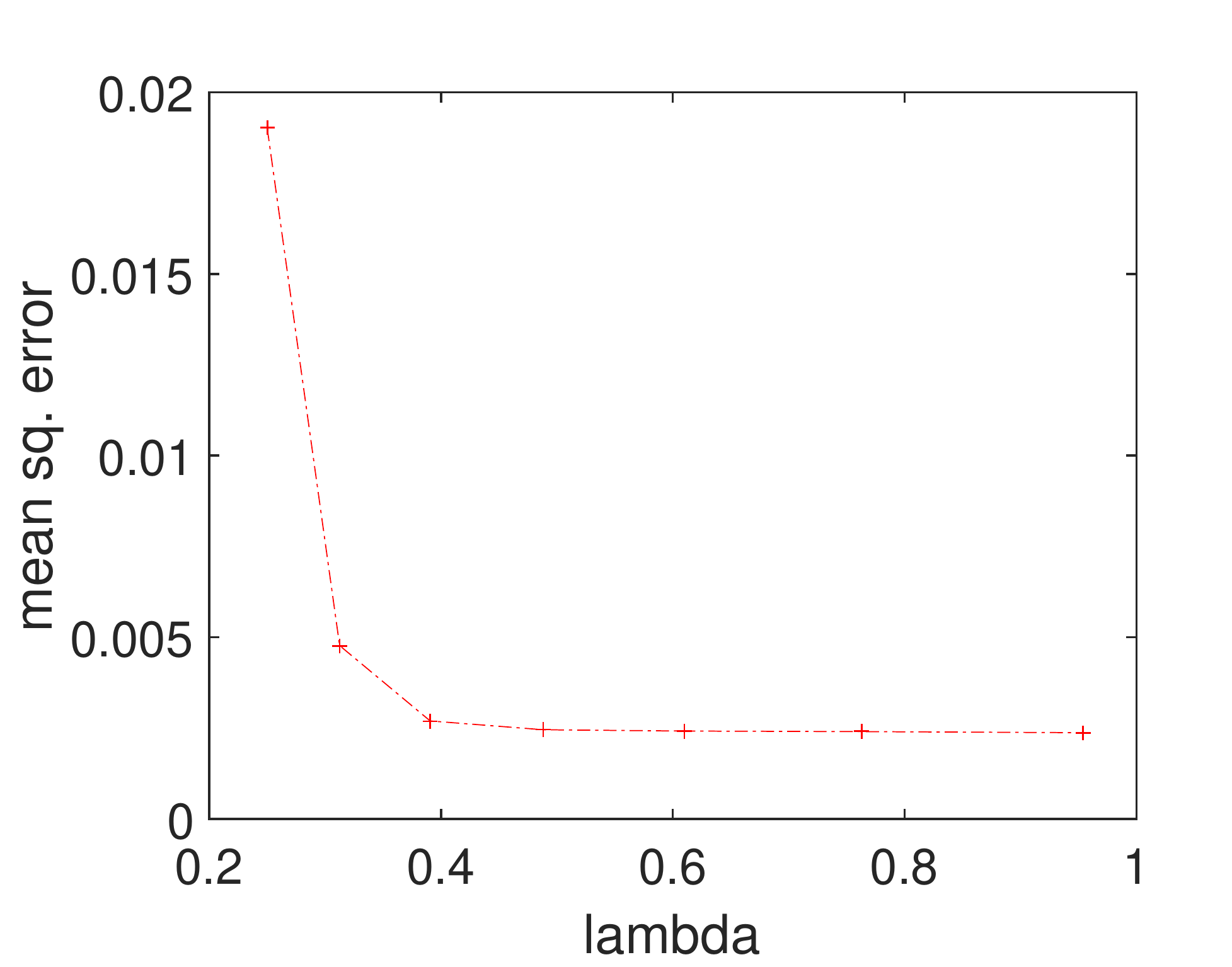}}
			\end{center}
			\caption{Consensus in optimization model and error in prediction (a) consensus over $\lambda$ (b) mean squared error over $\lambda$.}\label{fig:mseconsen}
		\end{figure*}

		In Figure~\ref{fig:mseconsen}, (a) we show how the consensus over the test set varies over different values of $\lambda$, the higher the value of the consensus indicates more nodes in the network are in sync. (b) Shows how the mean squared error (mse) varies with the $\lambda$. As seen, for the right range of $\lambda$, the $mse$ falls to the minimum as lambda slowly increases. This shows the prediction accuracy of the algorithm and its applicability to modelling the economic interrelationships of the underlying ride parameters; and robustness to early convergence resulting from the network consensus. Without the network consensus, the time taken to convergence would be cubic. Such accurate fare prediction can be used by the ride sharing application for efficient fare pricing by jointly factoring in that similar models of rides can be grouped together and shared. 
		
		\begin{figure}
			\begin{center}
				\subfloat[][]{\includegraphics[width=0.45\textwidth,clip=true,trim=0 0 0 0]{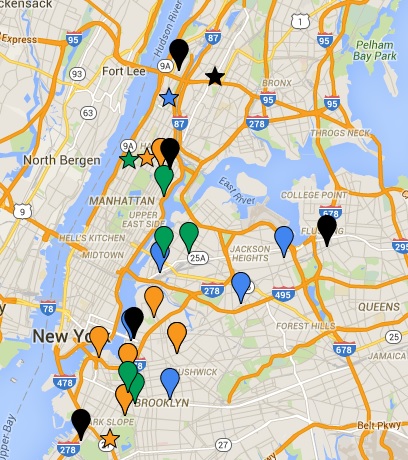}}\hspace{0.23cm}
				\subfloat[][]{\includegraphics[width=0.45\textwidth,clip=true,trim=0 0 0 30]{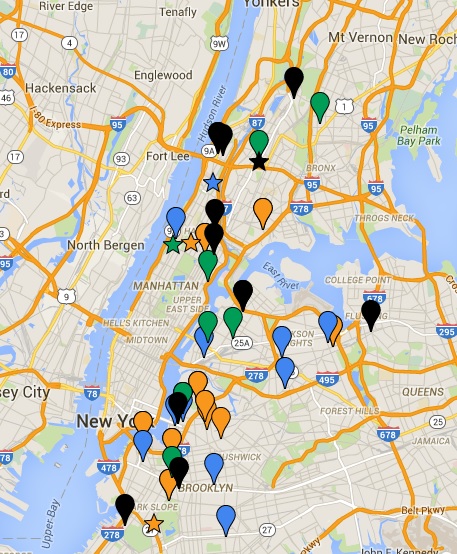}}
			\end{center}
			\caption{Clustering membership deduced for test data based on training clusters (a) $5$ neighbourhood structure (b) $10$ neighbourhood structure.}\label{fig:clus1}
		\end{figure}
		\subsection{Discussion}
\begin{figure}
			\begin{center}
				\subfloat[][]{\includegraphics[width=0.3\textwidth,clip=true,trim=0 0 0 0]{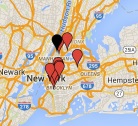}}\hspace{0.2cm}
				\subfloat[][]{\includegraphics[width=0.35\textwidth,clip=true,trim=0 0 0 0]{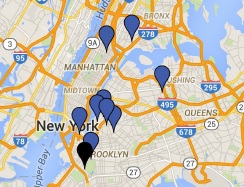}}
			\end{center}
			\caption{Deducing cluster membership for (a) Test trip node 77 indicated in black (b) Test trip node 193 indicated in black.}\label{fig:predindi}
		\end{figure}

\begin{figure*}[t]
			\begin{center}
				\includegraphics[width=0.6\textwidth,clip=true,trim=20 180 0 200]{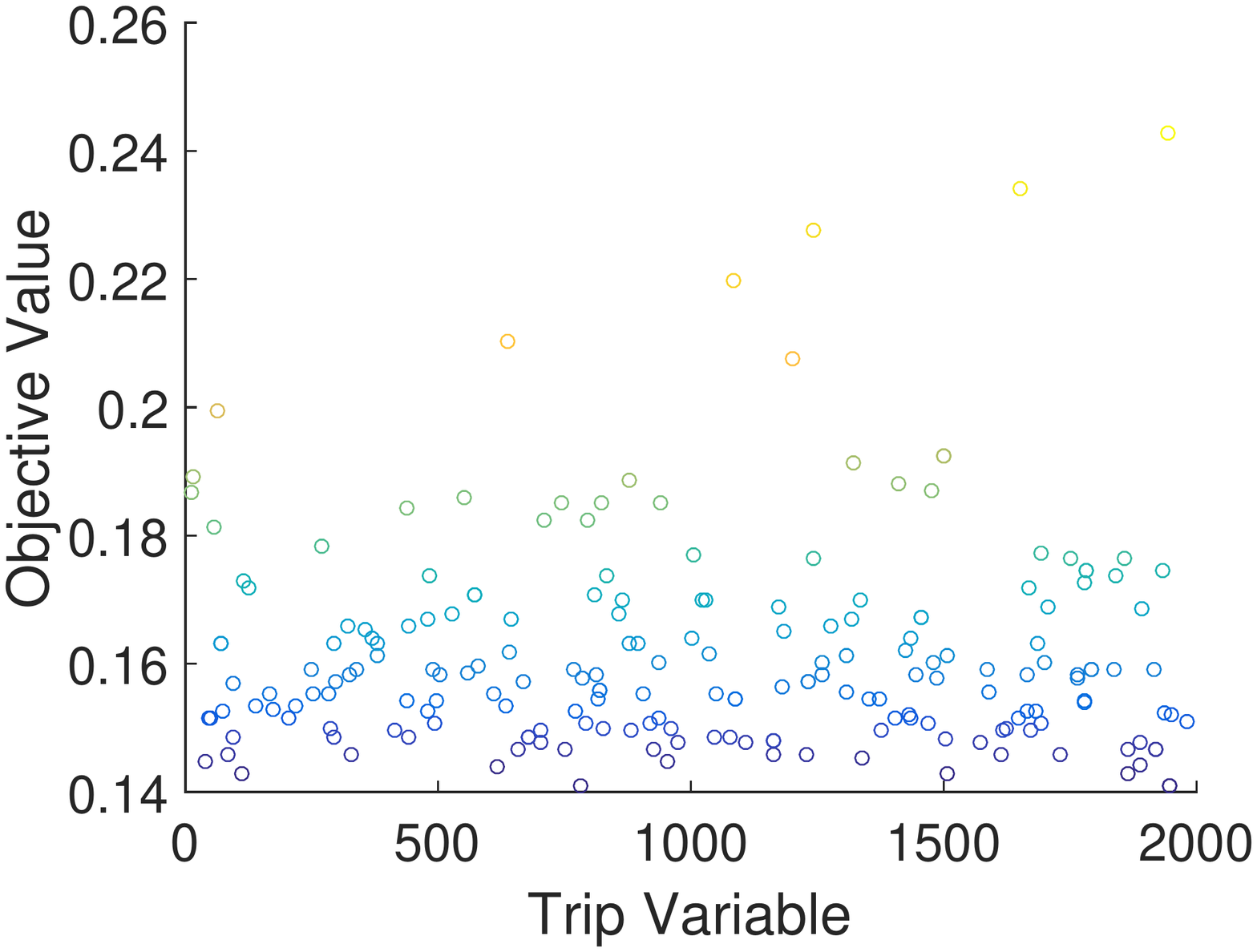}
			\end{center}
			\caption{Clustering based on Predicted Values on Training Set.}\label{fig:tripvars}
		\end{figure*}

\begin{figure}
			\begin{center}
				\subfloat[][]{\includegraphics[width=0.5\textwidth,clip=true,trim=70 210 92 200]{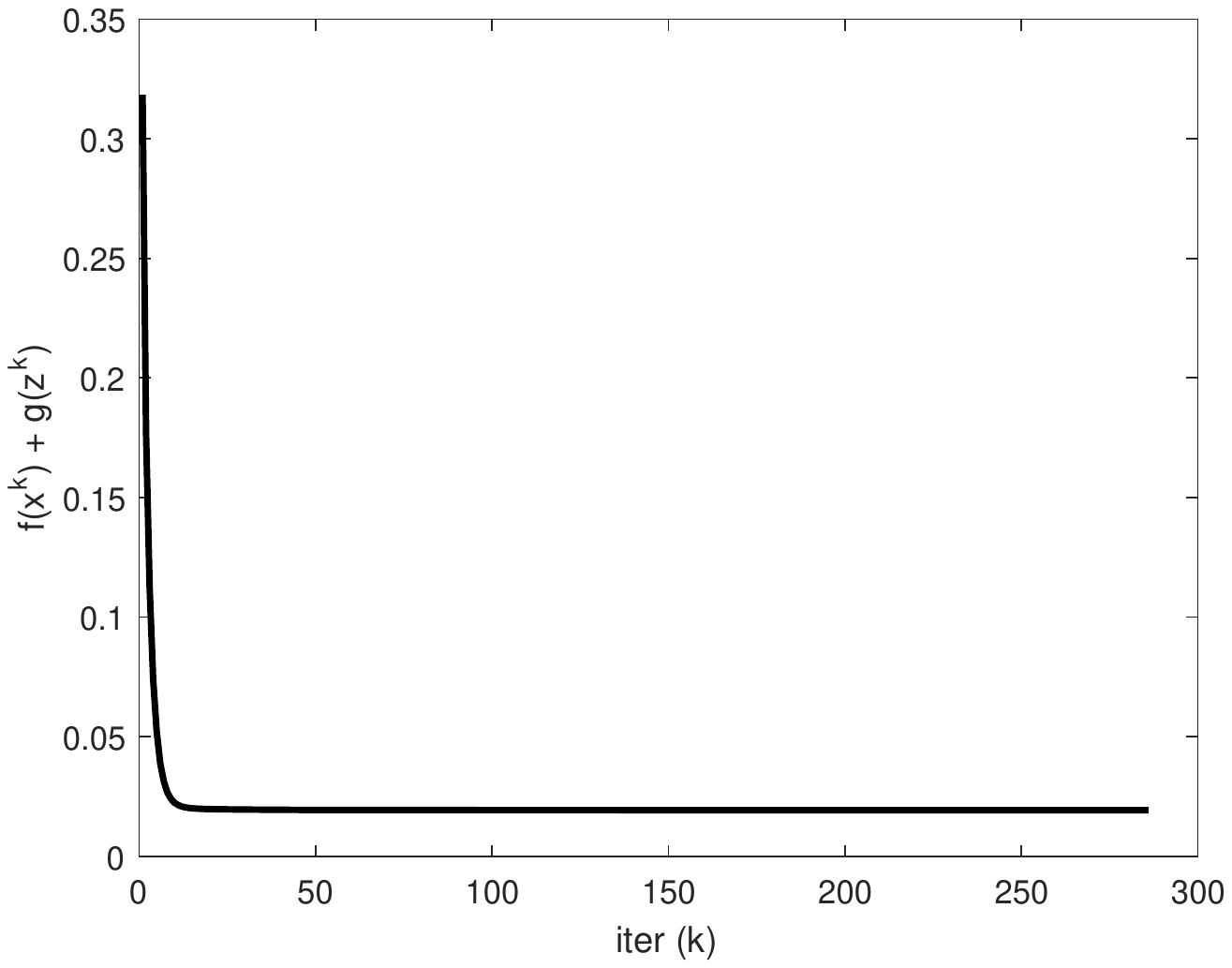}}
				\subfloat[][]{\includegraphics[width=0.5\textwidth,clip=true,trim=70 210 92 200]{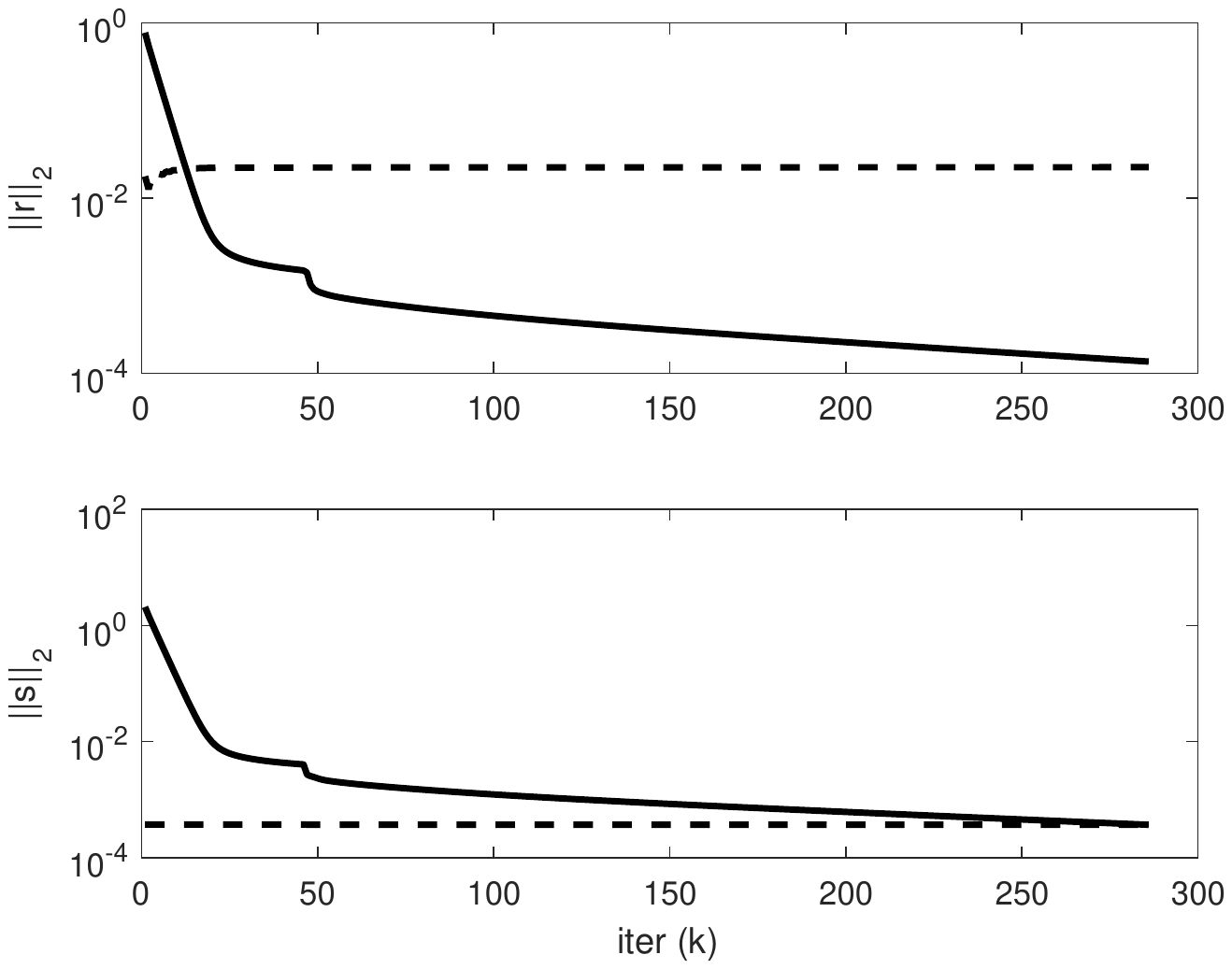}}
			\end{center}
			\caption{Objective values across iterations with ADMM Lasso.}\label{fig:lassobj}
		\end{figure}

		In Figure~\ref{fig:clus1}, we show the clustering of a portion of the test data set as performed by the algorithm. The test data point is indicated by asterisk markers, the colour of the marker is deduced by the algorithm which decides its cluster membership. There are four clusters (each indicated with a different colour) which the algorithm assigns to the test point such that its variables can be deduced based on the cluster to which it belongs, or is closest in terms of the similarity of their models. (a) uses a five neighbourhoods and (b) uses a ten neighbourhood structure resulting in more overlap. It is important to note the following observations. First, spatially distance rides can be grouped together if there is similarity in their model parameters along with the spatial closeness. This is an unique emergent property of this work in the context of ride sharing. Traditional ride sharing systems group rides that are only close geographically, but the method that we discuss is capable of doing both. Second, in traditional systems the grouping and the optimization are usually separate processes. Optimization decoupled from the network structure takes longer to converge, resulting in delay in responsiveness of serving rides. In Figure~\ref{fig:predindi}, we magnify the cluster that the algorithm decides the test point belongs to. The test point is indicated by the black markers. For the test point in (a) it belongs to the red cluster and the test point in (b) belongs to the blue cluster.	
		Figure~\ref{fig:tripvars} shows an alternative illustration of the clustering detected by the algorithm where the grouping location is not overlaid on the map and instead just shown on the basis of the predicted values. Similar models predict the similar value and the colour indicates the value ranges that the trip nodes belong to.
		The Figure~\ref{fig:lassobj} shows the vanilla Lasso prediction without the network data to validate the accuracy of prediction. Vanilla lasso converges to the minimum value of the objective function in terms of learning the model but is incapable of finding any cluster or groupings in the rides.
		\section{Conclusion}
		In this work, we make a novel connection between the ride sharing scenario and a scalable and robust optimization technique called Network Lasso optimization. We apply different well known techniques from the statistical regression analysis and machine learning paradigm to synthetic and real world data that encodes ride related attributes and variables, in order to perform joint optimization and clustering of similar rides that share similar models. To the best of our knowledge, this work is the first attempt in applying techniques that jointly optimizes and clusters rides in order to predict new ride sharing opportunities based on a network topology of rides. Rides that are similar in the modelling of their parameters get grouped together and hence can be shared. We evaluate the accuracy of the applications of Lasso, ADMM and Network Lasso on the synthetic dataset and real world dataset of green taxi trip records from the New York Taxi and Limousine Commission open data. We observe that vanilla ADMM Lasso achieves convergence within 50 iterations, but cannot sufficiently explore the network topology. Network Lasso however achieves an accuracy of 99.8 percent with a efficient clustering of 8 percent. We also notice that Lasso in itself ensures a sparsity and generalized model with automatic feature selection of 30 percent for rides with variables spanning 5000 dimensions in the synthesized dataset. We conclude that ADMM Network Lasso in particular is an efficient framework for large scale ride sharing systems that require distributed, scalable optimization with sufficient exploitation of the network topology of the similarity among rides for predicting new sharing opportunities. As, future work, we would like to extend the algorithms for prediction in less information environments where there is not enough training data due to lack of real life systems. 
		%
		%
		\bibliographystyle{plain}      
		\bibliography{references}   
		%
	\end{document}